%% file: quicksilver.tex
\documentclass[conference,compsoc]{IEEEtran}

\ifCLASSINFOpdf
  \usepackage[pdftex]{graphicx}
\else
\fi
\usepackage{amsmath}
\usepackage{booktabs}
\usepackage{multirow}
\usepackage{subcaption}
\usepackage[table]{xcolor}
\usepackage{xurl}

\definecolor{darkred}{HTML}{CB0000}
\definecolor{darkgreen}{HTML}{32CB00}

\ifCLASSOPTIONcompsoc
  \usepackage[nocompress]{cite}
\else
  \usepackage{cite}
\fi

\newcommand{\para}[1]{{\vspace{2pt} \noindent \textbf{#1}
    \hspace{6pt}}}

\newcommand{\swuedit}[1]{{\color{black} #1}}

\newcommand{\josephine}[1]{{\color{green} Josephine: }}

\newenvironment{packed_itemize}{
\begin{list}{\labelitemi}{\leftmargin=1em}
  \setlength{\itemsep}{1.5pt}
  \setlength{\parskip}{0pt}
  \setlength{\parsep}{0pt}
  \setlength{\headsep}{0pt}
  \setlength{\topskip}{0pt}
  \setlength{\topmargin}{0pt}
  \setlength{\topsep}{0pt}
  \setlength{\partopsep}{0pt}
}{\end{list}}

\begin{document}

\title{An Empirical Analysis of AI Slop in Music Streaming}

\author{
\IEEEauthorblockN{
Stanley Wu,
Josephine Passananti,
Viresh Mittal,
Wenxin Ding,
Haitao Zheng,
Ben Y. Zhao
}
\IEEEauthorblockA{
University of Chicago
}
}

\maketitle

\begin{abstract}
Generative AI models lower the bar for content creation, making it easy for any user to create professional-looking images, text and music with minimal effort. This has enabled a new cottage industry around creation of ``AI slop'' mass quantities of mediocre content produced to generate revenue, often through misrepresentation as human-authored content, or scams involving automated scripts and fake consumption.

While there are obvious parallels between the AI-slop industry and ``traditional'' email spam networks, it might be too early to determine if AI slop generation can grow into a similar self-sustaining industry. In this paper, we look specifically at the music industry, and explore the question: Can we prevent AI music slop from growing into a self-sustaining shadow industry?

To answer this question, we characterize the current state of AI slop in music, and its pipeline from generation, distribution, and consumption by users on streaming platforms. By examining growth and engagement on Spotify, we confirm that AI music exhibits AI slop characteristics: the overwhelming majority (93\%) of AI music receive few, if any listener plays, and are rarely recommended. AI musicians ``spray and pray,'' releasing large volumes of music across multiple genres in hopes of generating a hit. We also explore the AI slop pipeline by generating and publishing our own AI tracks onto streaming through 11 indie music distributors. We find distributors have inconsistent and largely unenforced policies on AI music, making it surprisingly easy to publish mass produced AI songs. Finally, we consider AI music detection, and find that current methods lack accuracy or robustness. As generation costs decrease, we believe slop generation in music will become self-sustainable, unless concrete steps are taken by the music industry. We consider and discuss potential mitigation methods based on our findings.

\end{abstract}

%
\IEEEpeerreviewmaketitle

\input{src/intro}
\input{src/back}
\input{src/methodology}

\input{src/crawl-eval}
\input{src/slop}
\input{src/ai-artist}

\input{src/detection}
\input{src/economics}
\input{src/conclusion}

\bibliographystyle{IEEEtran}
\bibliography{quicksilver}

\appendices
\input{src/appendix}

\end{document}

%% file: src/intro.tex
\section{Introduction}
\label{sec:intro}

\vspace{-0.07in}
Today's generative AI models lower the bar for content creation, making it
easy for any user to create professional-looking images, text and music with
minimal effort.  Not surprisingly, ``democratizing'' content creation to a
sequence of user prompts has led to a wave of low-cost, low-effort
content flooding different marketplaces previously dominated by
human creatives, now commonly referred to as ``AI
slop.\footnote{\url{https://en.wikipedia.org/wiki/AI_slop}}'' 

Today, AI slop seems to be everywhere~\cite{slopnymag}. It is overwhelming
Amazon's book marketplace~\cite{rollingstone}, digital
libraries~\cite{libraries}, small business platforms like
Etsy~\cite{etsy2,etsy}, and art stores, online platforms and art
conventions~\cite{artstores}.  Some of these AI works even target specific
works of individual human creatives, either by making copycat books of newly
released books~\cite{copybook}, mimicking music styles or voices of existing
musicians~\cite{copymusic,slopmusicimpersonation,slopmusicmimicry}, or
producing lookalike images based on individual human artworks~\cite{copyart}.

There are obvious parallels between the rapidly growing AI slop industry and
traditional email spam networks.  Despite global surveys that report 70\% of
consumers are uncomfortable with AI-generated content~\cite{variety}, bad
actors create AI slop because its production costs are so low, that any
minimal revenue will make slop production profitable. To be clear, not all AI
produced content is slop, but high volume, low effort slop content threatens
to drown out higher quality content, regardless of whether AI tools were
involved in its creation.

The impact of AI slop on the music industry is perhaps the most interesting,
both because of its large size (the global music streaming market generated
48.6 Billion USD in 2024~\cite{musicvalue}) and the late arrival of high
quality generative models for music like SUNO. Despite recent anecdotes about
large volumes of AI music on platforms like Spotify~\cite{slopmusicusers},
and scammers uploading AI music to inflate listen counts and steal royalty
payments~\cite{slopmusicscam}, there are no studies or empirical data to
answer critical questions about AI-generated music. For example, how much AI
music is there on today's music streaming platforms, and how much of it
exhibits properties of low-quality slop?  Is AI music slop growing into a
significant industry, and what if any mechanisms and controls can we bring to
prevent and throttle music slop?


To answer this question, we conduct a detailed empirical analysis of AI slop
on Spotify, the world's largest music streaming platform with 750 million
total users.  First, we analyze the prevalence of AI-generated music on
Spotify, using a metadata dataset of 256 million tracks accounting for 99\% of Spotify
tracks as of July 2025, and a recommendation graph containing a strongly
connected graph of 33 million tracks and 3.5 billion recommendation edges. 
Second, we explore the music slop generation process, by creating and
uploading our own music tracks to streaming platforms using independent
distributors. We learn that there are no hurdles or controls to throttle
music slop. Distributors are the primary ``curators'' of music, but
distributor policies against AI generated music are rarely enforced. Third,
we consider the critical missing piece of the curation process, the detection
of AI-generated music, and evaluate the detection accuracy and robustness of
automated detection tools. 

Our work produces a number of key findings.
\begin{packed_itemize}
\item We analyze the growth of AI music on the Spotify platform since 2024,
  and find significant signs of AI slop. Many AI musicians submit
  AI-generated songs across numerous genres, at a much higher volume than
  human musicians. Most AI music has very low engagement with listeners and
  recommendation systems.
\item Looking closer, a small subset ($<$3\%) of AI artists are creating
  large volumes of music at incredibly high rates, releasing more tracks than
  all other AI artists combined. On average, these AI slop artists generate
  lower quality music, and produce profitable ``hit'' songs at less than 1/3
  the frequency of other AI artists.

\item To better understand the pipeline of AI-generated music and potential
  throttle points for slop, we generate AI music tracks and submit them to 11
  independent music distributors (5 of whom state they will not distribute
  AI-generated music).  Regardless of their public policies on AI music,
  nearly all distributors approved our submissions and distributed them to
  streaming platforms. Only 2 denied our submissions for reasons unrelated to
  AI, and nearly all failed to identify our submissions as AI music.
  Clearly, the main hurdle preventing stakeholders (distributors, streaming
  services, users) from moderating AI music slop is the lack of an accurate
  way to identify AI music.
\item We evaluate automated detectors, and find the only accurate detectors
  are variants of a classifier from Deezer (French streaming platform). While 
  it is easily evaded by adversarial transformations like compression or
  pitch shift, we prove that it can be made robust through adversarial
  training.
\end{packed_itemize}

Our results highlight the risks that large volumes of low effort AI music
slop pose to the music industry, including human and benign AI musicians.
They also demonstrate that current stakeholders in the music ecosystem lack
the ability to identify or throttle music slop, an issue that must be
addressed to prevent music slop from growing into a much bigger problem.

%% file: src/back.tex
\section{Background}
\label{sec:back}

In this section, we provide necessary background and context for our study.
We start by defining AI slop, and discuss its similarities to email spam.  We
also give background information on the music distribution process, and
summarize current progress on AI music generators and detectors.

\subsection{AI Slop}
\label{subsec:ai-slop}

As described in the introduction, AI slop refers to the proliferation of
``lower'' quality, mass-produced content using generative AI
models. Typically, the motivation behind uploading AI slop is to generate
revenue via engagement/consumption achieved through the sheer volume of
uploaded AI content, or by manufactured fraudulent streams~\cite{slopmusicscam}.

\para{Similarities to email spam.}  General AI slop shares many
characteristics with email spam, including large volume, low-effort or cost,
and lower quality and user engagement. In email spam, seminal work like
Spamalytics~\cite{spamalytics} and Click
Trajectories~\cite{click-trajectories} shed light on the
quantitative impact of email spam through large-scale measurement studies,
often on live botnet systems. These works have provided significant insights
to security researchers on how to prevent spam and related fraud.

\begin{figure}[t]
    \centering
    \includegraphics[width=0.95\linewidth]{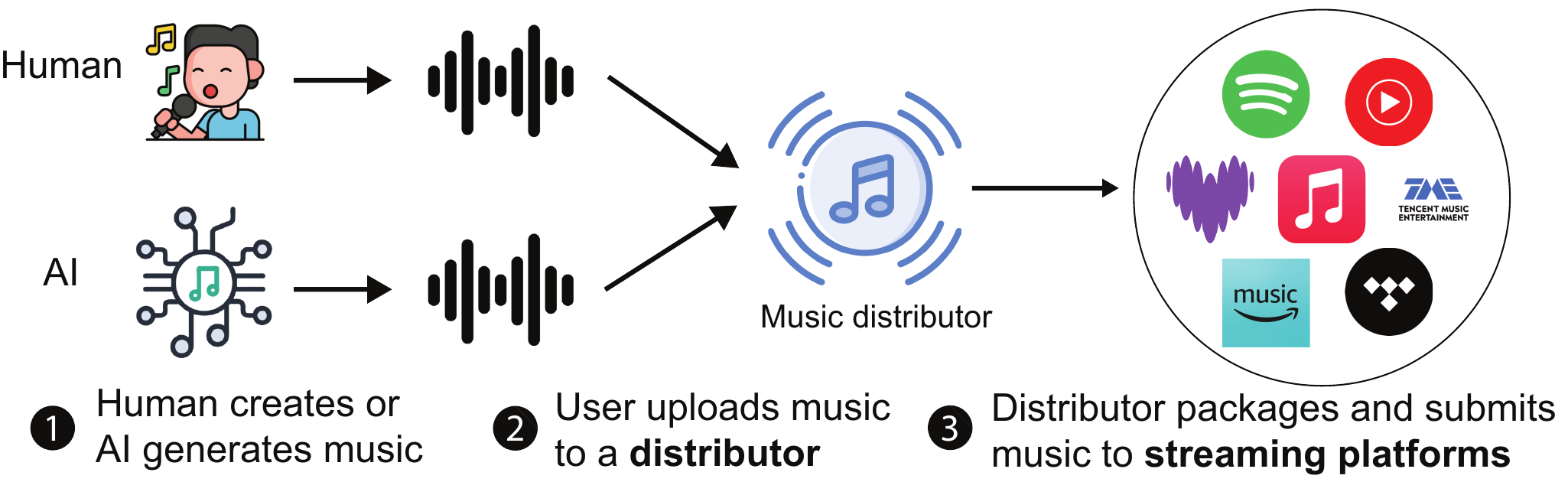}
    \caption{Full pipeline of how music (AI and human-created) is published to streaming platforms like Spotify.}
    \label{fig:ai-music-pipeline}
\end{figure}

\subsection{Music Streaming Ecosystem}
\label{subsec:distribution}
This work focuses on AI music slop and its delivery to streaming music
platforms. Below, we provide relevant information about the industry, and how
it has shaped our own study. To the best of our knowledge, no prior work has
attempted to measure AI music slop at scale.

\para{Streaming Platforms and Spotify.} Today, streaming \swuedit{is by far} the dominant method
for music consumption, and accounts for 84\% of revenue in the music
industry~\cite{streaming-industry}.  Popular streaming platforms include
Spotify, Apple Music, Amazon Music, Tencent Music, Deezer, Tidal, etc.
Spotify dominates the market by volume (31\%), more than twice that
of the second-largest platform, Tencent. It also has the largest listener
base, with over 761 million users~\cite{streaming-industry,spotifyusers}.

Musicians distribute their music across all streaming platforms because they earn
royalties from each platform~\cite{platform-exclusivity}, which results in music catalogs
across major streaming platforms overlapping significantly. As the largest
and dominant leader in music streaming, we focus our study on Spotify as the most
representative streaming platform.

\para{Uploading music via distributors.}  Today, major streaming platforms
like Spotify and Apple Music only accept music uploads through approved
distributors (see Figure~\ref{fig:ai-music-pipeline}).  Distributors act as
``middle-men'' between musicians and streaming services, managing tasks such
as audio formatting, metadata packaging, and collecting
royalties~\cite{music-distribution}.  In return, distributors are paid either
by upfront fees, or by taking a percentage of the royalties earned.  When
musicians upload tracks to a distributor, they specify which streaming
platforms to target. Distributors also play an important role in validating
music for copyright violations, duplicate submissions, and optionally apply
their own policies on distribution of AI music~\cite{distributorsai}. Later in
\S\ref{sec:distribution}, we explore the AI music distribution process by
tracking the flow of \swuedit{our own} AI tracks to streamers via 11 music
distributors.

Our study considers music distributed from a wide range of
distributors. Major record labels serve some of the most popular musicians,
but only a small portion of all music by volume. Studies show that the large
majority of music (96\%) is distributed outside the major labels and their
subsidiaries~\cite{small-labels}.  Thus, our study focuses on music
distributors (rather than large record labels) as the backbone of the music
ecosystem.

\para{Distributors' policies on AI music.}  Distributors vary in their policies
on AI-generated music. Only a few explicitly prohibit AI music.  However,
many have proactively deployed AI-detection systems to enable quality control
and enforce copyright compliance.  For instance, two distributors, Amuse and
Ditto, collaborate with SubmitHub~\cite{submithub-workwith}, a commercial
AI music detection service.

\para{Streaming services' policies on AI music.}  Notably, Deezer is the only
streaming platform that publicly labels the AI music they
host~\cite{deezerailabel-june,deezerforbes} using an in-house
detector. According to their \swuedit{latest report}, 44\% of their daily incoming music is
fully AI-generated, and 85\% of AI tracks' ``streams'' are
fraudulent~\cite{deezerailabel-april}. Deezer does not include AI tracks
in their music recommendations~\cite{deezerforbes}. In contrast, Apple Music recently
announced AI transparency tags, but rely entirely on music distributors as
their mechanism for identifying AI-generated music~\cite{applemusic}.

Spotify has also taken action against AI-generated music spam, but has no
plans to publicly label and/or exclude from recommendations.  Spotify did
announce in 2025 that the platform has already removed over 75 million
`spammy' AI tracks, and that they are working on AI disclosure in
collaboration with 15 distributors~\cite{spotify-remove-tracks}. Spotify also
recently announced the rollout of an artist verification
badge~\cite{spotify-verification-badge} to distinguish real artist profiles
from AI personas.

\subsection{Generators and Detectors}
\label{subsec:generators}
\para{AI music generators.}  Modern AI music generators leverage either
autoregressive~\cite{dhariwal2020jukebox,copet2023simple} or latent diffusion
models~\cite{forsgren2022riffusion, evans2025stable, diffrhythm, acestep}.
In this study, we focus on commercially viable music for mainstream streaming
platforms. Thus we only consider models that produce full-length, lyrical and
instrumental music, and exclude models that synthesize short beats or
melodies which are mainly used to assist artists during the creative
process. These full-length models allow users to generate full-fledged songs
with minimal user input, and so are best suited for the generation of AI
slop.

Very few open-source models are capable of generating full-length songs.
We identified only two:
DiffRhythm~\cite{diffrhythm} and ACE-Step~\cite{acestep}.  Both are
diffusion-based, run on consumer-grade GPUs, and generate complete songs
in under a minute.  Earlier models (e.g. SongGen~\cite{liu2025songgen},
Stable Audio Open~\cite{evans2025stable}, Musika~\cite{pasini2022musika},
Riffusion~\cite{forsgren2022riffusion}) fail to meet these requirements.

In contrast, the number of commercial AI music generators has grown rapidly,
with five generators leading the space~\cite{ai-generator-report}.
Suno~\cite{suno} and Udio~\cite{udio} can generate full-length songs, while
Boomy~\cite{boomy}, AIVA~\cite{aiva}, and Soundraw~\cite{soundraw} focus on
beat sampling, stemming, and instrumentals, with limited functionality to
generate full-length songs.  For Suno and Udio, an annual subscription of
\$100 USD allows users to generate roughly 500 songs per month ($<$0.02\$ per
song), each taking roughly a minute to generate.

Our study uses all four generators capable of generating full-length songs:
DiffRhythm and ACE-Step (open-source), and Suno and Udio (closed-source
commercial).  Each allows fast and cheap generation of AI music, a key
factor needed to produce AI music slop at scale~\cite{slopmusicusers,slopmusicumg}.

\para{AI Music Detectors.}
\label{subsec:detectors}
Existing literature on AI music detection has
focused on a binary classification of spectrogram data using convolutional
neural networks~\cite{cnn-reconstruction,fourier-peaks,li2024audio,quicksilver} and
transformer-based architectures to improve feature
representation~\cite{sonics,ai-arms-race}. However, their performance depends
heavily on the audio datasets used for training. As we show later in
\S\ref{sec:ai-music-detection}, current detectors vary significantly in their
detection performance.

There are also commercial detectors only offering paid online
services~\cite{submithub,hive-music,matchtune,beatstorapon}.  They 
are closed-source, with little to no information on the underlying detection
method and training data. Only SubmitHub~\cite{submithub} disclosed that it
leverages techniques from an open-source detector~\cite{sonics}. At the scale
of our large datasets, we lack the resources to evaluate these commercial
detectors, except indirectly in the case of SubmitHub, which is used by two
distributors in our experiments, Amuse and Ditto~\cite{submithub-workwith}.

%% file: src/methodology.tex
\section{Goals and Methodology}
\label{sec:methodology}

Our study seeks to understand, at scale, the current and potential future
impacts of AI-generated music slop on the music streaming platforms. We are
particularly interested in quantifying its prevalence and popularity on the
largest streaming service (Spotify), and the potential challenges faced by
the current music distribution and streaming ecosystem in throttling AI music
slop.  Here, we describe the key questions and methodology used in our study.

\subsection{Key Questions}
\label{subsec:questions}

Our study is focused around several key questions.

  \para{Q1} How much AI music is available on Spotify and how fast is it growing?

  \para{Q2} How much of AI music is slop? How well does it engage with
  recommendation systems and users, and does it generate significant revenue?

  \para{Q3} What is the process for distributing AI music? Does it offer
  potential mechanisms to throttle or limit the growth of music slop?

  \para{Q4} What potential challenges, if any, does the music industry face
  in future moderation of music slop?

\subsection{Methodology}
We answer the above questions through a large measurement and experimental
study focusing on Spotify, the largest global music streaming platform (see
\S\ref{subsec:distribution}). Getting access to large-scale metadata on music
tracks is challenging. Existing public datasets are either outdated and do
not include data since emergence of generative AI (FMA~\cite{fma_dataset}),
or are users contributed datasets that capture only a sparse subsample of
today's music (MusicBrainz~\cite{stutzbach2011musicbrainz},
last.fm~\cite{lastfm}).

Instead, we use a recently released dataset of Spotify metadata, along with a
dataset of Spotify recommendation links we crawled (more details next in
\S\ref{subsec:datasets}). We label AI-generated tracks on Spotify by
cross-referencing tracks with labels of AI music on Deezer. We use the
resulting dataset to analyze the growth of AI music, the behavior of AI
musicians, engagement of AI music with users and recommendations, and
distinguish AI music slop from non-slop AI music.

\para{Real-world experiments on publishing AI music.} To obtain a deeper
understanding of how AI music \swuedit{slop} enters the music streaming
pipeline and how to throttle it, we run experiments where we generate and
publish AI music through 11 music distributors. We observe distributors'
current policies on AI music, their enforcement of those policies, and the
resulting role they play in the growth of AI music and slop today.  The
results also shed light on the potential role of AI detectors and economic
factors might play moving forward.

\para{Understanding the challenges of detecting AI-generated music.}  We
identify the challenge of detecting AI music as a key hurdle to the
moderation of music slop moving forward, and evaluate the efficacy of
algorithmic-based detectors. We test the
reliability of 4 automated AI music detectors under both normal and
adversarial conditions, and demonstrate how to improve their robustness under
attack.

%% file: src/crawl-eval.tex
\section{Initial Analysis of AI Music}
\label{sec:crawl-eval}

In this section, we describe our Spotify music datasets, including
how we labeled AI tracks, and initial analysis on the growth of AI music. This
section seeks to answer question (Q1) in \S\ref{subsec:questions}.

\subsection{Gathering and Labeling Spotify Datasets}
\label{subsec:datasets}
We now describe details of the two Spotify datasets used in our study, and
how we label AI-generated music in each.

\para{Spotify's Music Metadata adapted from Anna's Archive (SMMA).} In
December 2025, Anna's Archive released a dataset~\cite{annas-archive-spotify}
containing metadata of 256 million song tracks, claiming to cover 99\% of
Spotify's music catalog as of July 2025.  The dataset contains metadata for
each track, including its Spotify ID, International Standard Recording Code
(ISRC), total play count, release date, artist, and album name.  We
downloaded this metadata dataset without accessing any audio files.  Using
the ISRC (a universal identifier for recorded music) to remove duplicated
entries, we identified 185 million unique tracks, of which 53 million were
released after 2024. We used a number of heuristics to test the dataset's
claims, All test results were consistent with the dataset claims. We refer to
this dataset as SMMA.

\para{Spotify's Core Recommendation Graph (SCRG).} The SMMA dataset does not
capture relationships between tracks or how users might discover music
through their listening patterns. For this, we collected a separate dataset
of recommendation links from Spotify's recommendation network using their
recommendation API. This API is publicly accessible without account
credentials or authentication\footnote{We discuss our use of this API,
  ethical considerations, and disclosure in the Appendix~\ref{app:ethics}}. Using this API, we obtained the
largest strongly connected component of Spotify's recommendation network up
to May 2026 (details on our crawl are listed in
Appendix~\ref{app:spotify-webcrawl}).  In this graph, tracks are represented
as nodes, and their recommendation relationships are defined by directed
edges.  In total, we have 33 million tracks and 3.5 billion edges, as well as
the metadata for each track (similar to SMMA).  We refer to this dataset as
SCRG.  Notably, 99.4\% of tracks in SCRG are also in SMMA. 0.28\% (97k
tracks) were posted after July 2025, thus were not included in SMMA.  We do
not know the origin of the remaining 120k tracks, though they represent a
tiny (0.05\%) portion of the Spotify catalog.

Table~\ref{tab:dataset-overlap} summarizes the two datasets. In the rest of
the paper, unless otherwise stated, we use the combined dataset to analyze the
entirety of Spotify's music catalog.  

\begin{table}[t]
  \centering
    \resizebox{0.3\textwidth}{!}{
        \begin{tabular}{@{}ccc@{}}
        \toprule
        \textbf{Dataset} & \textbf{\begin{tabular}[c]{@{}c@{}}\# of \\ tracks\end{tabular}} & \textbf{\begin{tabular}[c]{@{}c@{}}\# of tracks posted\\ on/after 2024\end{tabular}} \\ \midrule
        SMMA & 185,361,997 & 52,915,713 \\
        SCRG & 33,512,395 & 6,804,105 \\
        Combined & 185,578,495 & 53,042,110 \\ \midrule
        w/ label & - & 40,748,961 \\ \bottomrule
        \end{tabular}
    }
    \caption{Spotify tracks captured by each of the two datasets, including
      tracks posted since 2024 since Suno and Udio arrived in early 2024.
      Cross-referencing tracks on Deezer produced labels for 40.7 M tracks.}
  \label{tab:dataset-overlap}
\end{table}

\begin{figure*}[t]
  \centering
  \begin{subfigure}[b]{0.42\linewidth}
    \caption{SMMA}
      \centering
    \includegraphics[width=\linewidth]{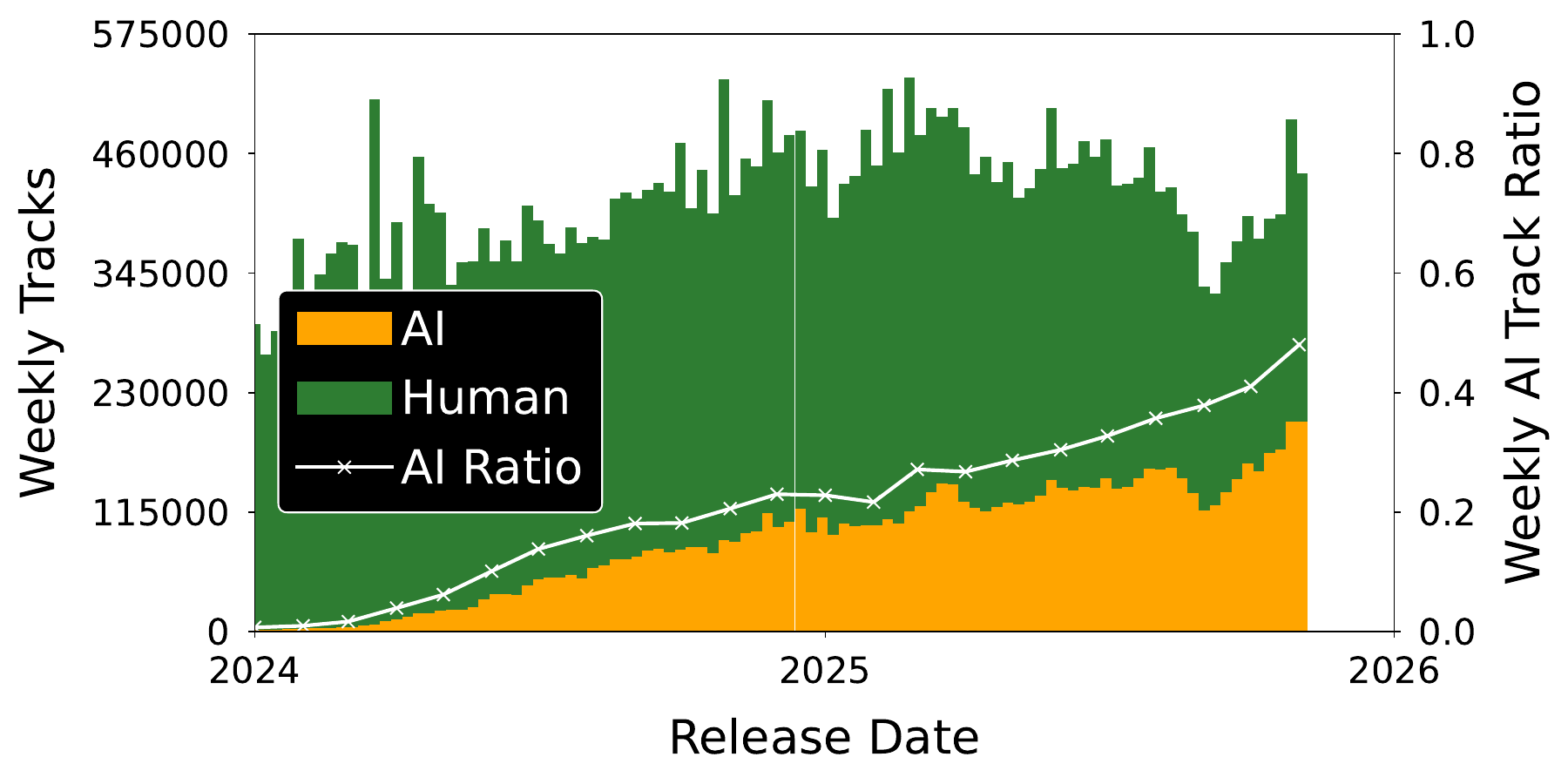}
    \label{fig:ai-music-over-time-anna}
  \end{subfigure}
  \hspace{0.2in}
  \begin{subfigure}[b]{0.42\linewidth}
    \caption{SCRG}
      \centering
    \includegraphics[width=\linewidth]{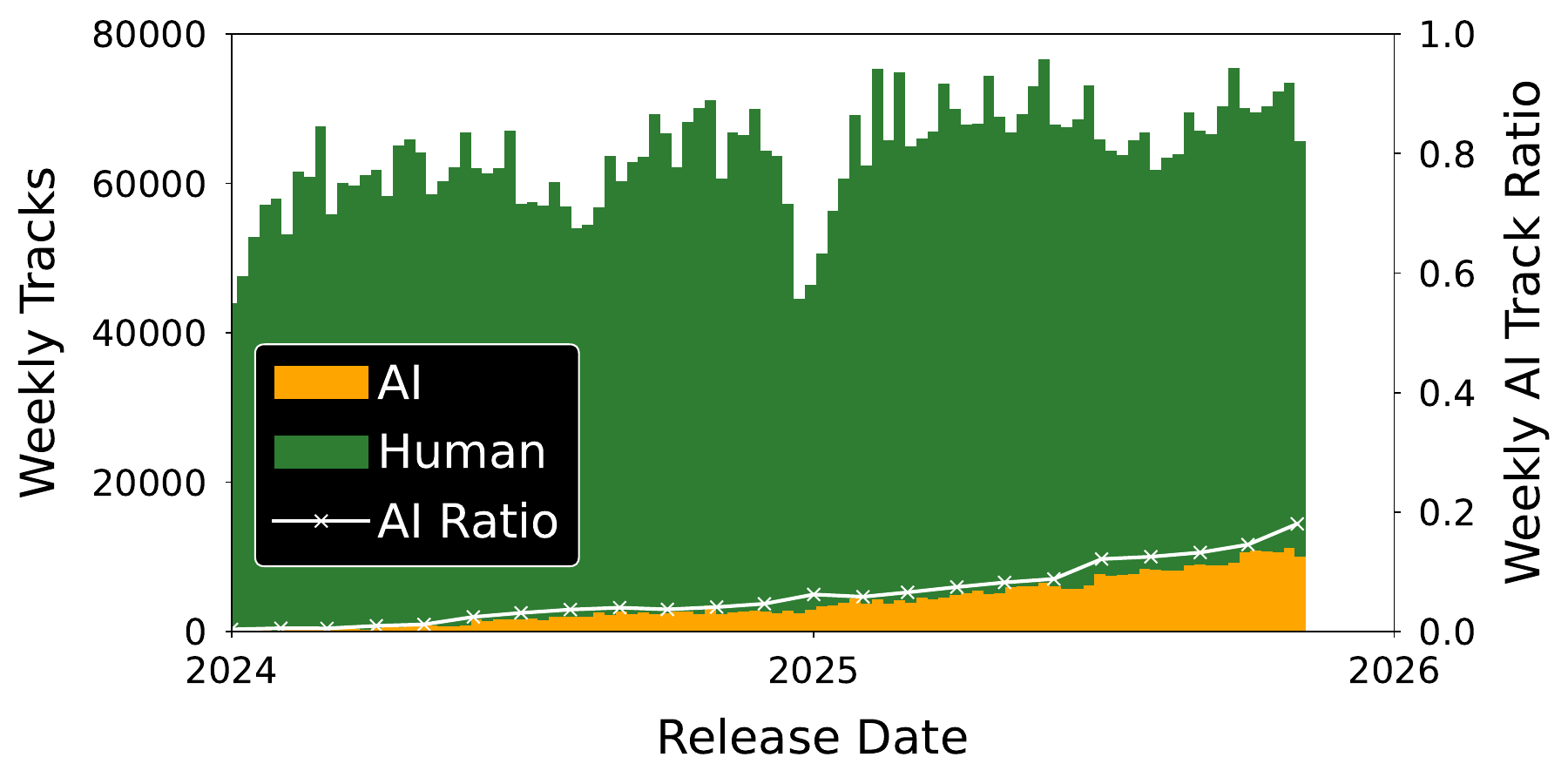}
    \label{fig:ai-music-over-time-scc}
  \end{subfigure}
  \caption{Stacked volumes of weekly releases of AI-generated and human-made
    tracks for (a) SMMA (overall music catalog) and (b) SCRG (strongly
    connected recommendation graph).  Both datasets
    show strong growth of AI music since 2024.}
  \label{fig:ai-music-over-time}
\end{figure*}

\para{Labeling AI music using Deezer labels.}  Neither dataset includes
labels indicating whether a track is human-made or AI-generated. Testing
each track using AI music detectors would have required us to download music,
likely violating artists' copyright in the process. 
Instead, we leveraged existing labels of AI music from Deezer. Deezer
is the only streaming platform that explicitly labels AI music, and does so
using a proprietary detector developed by their research team. As we show
later in Section~\ref{subsec:ml-detection}, Deezer's detector is also the most
accurate AI detector in our tests.  We identify and label AI music tracks in
our two datasets by cross-referencing its corresponding label on Deezer\footnote{
  \swuedit{Deezer's detector is designed to identify {\em fully AI}-generated tracks, but 
  Deezer only labels {\em albums} as AI if it contains {\em any} AI tracks.
  After studying key artists who produce fully AI-generated songs, such as 
  Timbaland~\cite{timbaland-ai}, Grimes~\cite{popular-artists-ai}, and The Velvet Sundown~\cite{velvet-sundown-ai}, we found they use AI consistently across all their productions, and
  do not mix AI-generated tracks with human-made songs. As such, we decided to label all tracks in an album the same based on that album's label on Deezer.}}.

We limit our matching-based labeling to tracks released on/after January
2024 \swuedit{in order to capture} the arrival of the advanced commercial AI music
generation models: Suno and Udio. Filtering by this timeframe, we have 52.9
million unique tracks introduced to Spotify after January 2024. We were able
to definitively match more than 77\% of these tracks with Deezer (40.7
million), where each track is labeled as AI-generated or not (details on how
we performed matching in Appendix~\ref{app:deezer-match-algo}). Looking closer
at the remaining 12.3M tracks not found on Deezer, we find that most were tracks 
from countries where Deezer was not available or blocked,
e.g., India, Belarus, Russia, and Sudan. We believe the 40.7M matched songs
represent a nearly complete set of music tracks in Western
countries released since 2024, and use it as the primary target of our
subsequent analysis.

\subsection{Growth and Popularity of AI-generated Music}
\label{subsec:ai-music-growth}

We start with a longitudinal analysis on the growth of AI music on
Spotify, using the labeled data of the two Spotify datasets (SMMA and
SCRG). As discussed earlier, SMMA represents Spotify's overall music
catalog while SCRG captures its tracks strongly connected by recommendation links. 

\para{Significant AI growth since 2024.} Figure~\ref{fig:ai-music-over-time}
plots for both datasets the number of AI tracks released on Spotify every week since
2024, stacked with human-made tracks in the same time frame.
Both indicate a steady growth of AI music. Note that our observed counts for
AI-generated music are likely a lower bound of the actual number of
AI-generated tracks that arrived on Spotify, since we cannot observe the
reportedly 75 million AI-generated tracks removed by
Spotify~\cite{spotify-remove-tracks}. 

In terms of the overall music catalog (SMMA), AI-generated music is growing
at a tremendous speed. Less than 1\% of weekly new releases were AI-generated
in January 2024, a figure that has increased beyond 40\% by Nov. 3rd 2025,
the last full week in which we have data
(Figure~\ref{fig:ai-music-over-time-anna}). In that last week, 202,046 AI
tracks were released, compared to 238,646 human tracks (60,022 tracks remain
unclassified, since they were not found on Deezer).  At that pace, AI tracks
has likely already surpassed human-created music for new releases in 2026.
Overall, AI-generated music already accounts for 5.1\% of the entire music
catalog, despite only being around for the last 2 years.

The growth of AI music is much more muted on SCRG, the strongly connected
recommendation graph (see Figure~\ref{fig:ai-music-over-time-scc}). There,
AI-generated music accounts for only 1.2\% of the SCRG tracks. Since tracks
on SCRG capture what the streaming platform predicts listeners will prefer
(which triggered the platform recommendations), this observation suggests
that listeners still strongly prefer human-made music. We dive deeper into
the user engagement with AI tracks in the next section.

%% file: src/slop.tex
\section{Analysis of AI Music Slop}
Next, we dive deeper into our
data to answer Q2 from \S\ref{subsec:questions}: how much of AI-music today
exhibits properties of AI slop, and how does it engage with users,
recommendation systems, and monetization.

We summarize our key observations below:

\begin{packed_itemize}
  \item AI music receives much lower engagement (play count) by users
    compared to non-AI-created music. 
  \item Significant portion of AI musicians upload a larger volume of music across many
    different genres, indicative of the expected low-effort, ``spray and
    pray'' strategy consistent with slop generation.
  \item AI music are more likely to recommend each other (like spam/engagement farms)
  \item AI music production is dominated by a small minority of slop
    producers, whose tracks overwhelm both human musicians and benign AI musicians.
\end{packed_itemize}

\subsection{Listener Engagement with AI Music}
\label{subsec:ai-slop-engagement}

We begin with a closer look at user engagement for AI-generated
music versus human-created music. 

\para{AI music playcounts.}  We first analyze the raw cumulative play counts
for both AI and human-made music. Our results found that 93\% of AI music
tracks have received less than 1,000 plays since its release, the minimum
threshold set by Spotify for
monetization~\cite{spotify-monetization-eligibility}.
In contrast, only 64\% of human-made tracks failed to reach this engagement
threshold during the same period.

Since AI-generated music is a more recent and growing phenomenon, a simple
lifetime playcount likely biases numbers against AI-generated tracks. Thus we
take a look at a fixed set of songs (AI-generated and human-generated) and
look at their playcount during a 5 month period, from January 1 to May 31, 2026.
We note that Spotify only provides play count information for tracks
with at least 1,000 plays, and does not provide exact play count below that
value since 1k plays is the threshold for monetization. These tracks 
are labeled as having ``negligible plays.''

\begin{table}[t]
  \resizebox{0.50\textwidth}{!}{
    \begin{tabular}{cccccc}
      \toprule
      \multirow{2}{*}{}     & \multicolumn{5}{c}{\textbf{\% of tracks based on \# of plays from Jan. to May 2026}}   \\ \cmidrule(l){2-6}
                            & negligible plays & \multicolumn{1}{l}{$\geq$10} & $\geq$1k & $\geq$100k & $\geq$ 10mil \\ \midrule
      \textbf{AI-generated} & 92.7\%           & 6.9\%                        & 4.7\%    & 0.1\%      & 0.0002\%     \\
      \textbf{Human-made}   & 67.5\%           & 30.5\%                       & 10.5\%   & 1.0\%      & 0.02\%       \\ \bottomrule
    \end{tabular}
    }
    \caption{We report the ratio of tracks under
      different buckets of play count from Jan to May 2026, separated by AI and human-made music.}
  \label{tab:tracks-playcount-cdfs}
\end{table}

Table~\ref{tab:tracks-playcount-cdfs} lists the statistical breakdown of
tracks by their play counts during the first 5 months of 2026, separately for
AI-generated and human-made music. These results show the same trends as our
aggregate play counts: human-made tracks are much more popular than AI tracks
for a static set across the same 5 month timeframe.  In fact, the
overwhelming majority of AI tracks (92.7\%) received negligible ($<$1000
total) plays during this entire 5 month period. In comparison, only 67.5\% of
human-made tracks received negligible plays. Human-made music produced
dramatically higher engagement, particularly in the highly popular ($>$200k)
or ``smash hit'' range ($>$10 million) of play counts.

\para{Repeated engagement.}  In the music industry, listeners tend to return
to artists they have already discovered whose content they like. Thus,
musicians who release popular songs are more likely to experience higher play
counts for subsequent releases. Thus we can use this spillover effect as a
secondary measure of user engagement.

We find that this repeated engagement is much weaker for AI-generated music.
We measure this pattern by comparing the per-track play counts of a
musician's tracks before and after the release of their first top hit. We
define a top hit as a song whose total play count falls within the top 1
percentile of all songs in that category (either AI-generated or
human-created).  For each artist with at least one top hit, we divide their
songs into two groups: those released before the first top hit, and those
after. For tracks in each group, we measure the average number of plays per
day since its release, and compare the median value for tracks in group 2 to
group 1.

For human musicians, the median number of plays per day increases by 16x for
tracks after the first top hit.  Releasing a top hit produces a huge boost to
play count of subsequent releases.  For AI music, the boost is much weaker,
only 5x. That is, users are less likely to follow AI musicians to find their
subsequent releases. Overall, it is clear that users engage with AI-generated
music very differently than they do with human-made music.

\subsection{Many AI Artists Exhibit Spammer Behavior}
\label{subsec:ai-music-properties}

Next, we turn our attention to the behavior of the artists who create the
music. Given our general definition of slop as ``low effort'' ``high volume''
content, observing the behavior of creators is perhaps an even better way
to identify real slop than content quality.

First, we classify a ``human musician'' as someone who has never released any
AI-generated music, and an ``AI musician'' as someone who has released at least
one AI-generated track.  Intuitively, there are three relevant types of AI
musicians: {\em AI-only} (who only release AI-generated tracks), {\em
  switched} (former human musicians who then switched to
AI-only), and {\em mixed} (who release both human-made and AI-generated
tracks).

Table~\ref{tab:musicians} shows the number of human musicians compared to the
three types of AI musicians found in our combined Spotify dataset.  One
interesting (but perhaps unsurprising) finding is that AI musicians and human
musicians tend to be very disparate groups. The majority of AI musicians
(79\%) are AI only, i.e. they have never released music before generative
AI. There is a {\em very} small group of human musicians who switched 
to generative AI (less than 0.27\% of all human artists).

\begin{table}[t] 
  \centering 
  \resizebox{0.45\textwidth}{!}{
    \begin{tabular}{@{}cccc@{}}
      \toprule
      \multirow{2}{*}{\textbf{\# of human musicians}} & \multicolumn{3}{c}{\textbf{\# of AI musicians}} \\ \cmidrule(l){2-4} 
                       & \textbf{AI-only} & \textbf{switched} & \textbf{mixed} \\ \midrule
      13,320,045 & 266,090 & 35,902 & 69,328 \\ \bottomrule
    \end{tabular}
  }
  \caption{The number of human musicians and the three types of AI
    musicians found in our
    combined Spotify dataset.}\label{tab:musicians} 
\end{table} 
 
Let's now look closer at the behaviors of AI musicians vs. human musicians.

\para{Track count and upload frequency.} \swuedit{We first study music uploads.  Since
2024, each AI musician has uploaded 27 AI-generated tracks on average, double
the average volume of recently active human artist (13 tracks). 
When we normalized the number
of tracks released per month, human musicians average only 1 track, while AI
musicians average 5. Additionally, the average
gap between releases is a mere 16 days for AI musicians and 50+ days for
human musicians.  AI musicians are clearly producing music at a rate far
faster than human musicians. In the most extreme case, we even
found an AI artist who has uploaded over 60,000 AI tracks\footnote{
  \url{https://open.spotify.com/artist/2DsZ91fHYviRw9sTGgJdbv}
}. Later in \S\ref{subsec:separating-slop}, we look to see how these
characteristics allow us to identify slop from benign AI music.}

\para{Genres.} Human musicians often focus on a single genre in their
releases, as musicians tend to self identify with a single genre (with rare
exceptions like cross over artists who straddle two genres). In contrast,
producing high quality music in more than two genres seems extremely unusual,
even for musicians who are being assisted with AI tools. But when we examine
the genre diversity of AI musicians, we find that they cover much more
distinct genres than human musicians.  The most diverse (top 1\%) AI
musicians have each covered more than 12 different genres (some even reach 57
genres). 

We believe this is perhaps the strongest indicator of music slop in
AI-generated music. While music generators can produce full-length
tracks across many genres, it is difficult to imagine any artist being
familiar with numerous genres, much less capable of producing high quality
content across all of them. Instead, a much more plausible explanation for
the wide diversity of genres is that some AI musicians are focusing on
casting a wide net in hopes of producing a hit, a strategy often referred to
as the ``spray and pray'' approach. This is the type of spammy behavior often
associated with large-scale phishing or email spam
campaigns~\cite{spraypray}.

\subsection{AI Music and Recommendations}
\label{subsec:ai-music-accessibility}

On a streaming platform like Spotify, the key feature driving music discovery
is its recommendation system~\cite{spotify-recommendations}. Next, we study
how AI music is positioned within recommendation networks using the SCRG
dataset. SCRG represents the largest strongly connected component of
Spotify's recommendation network up to November 2025, with 33 million tracks
and 3.5 billion edges. Here each edge is directed: an edge $a \rightarrow b$
means that after a user listens to track $a$, Spotify had recommended the
user to listen to track $b$.  For simplicity, we call {\bf $a$ recommended
  $b$} if $a \rightarrow b$ appears in SCRG.

\para{Recommending AI music.}  AI-generated tracks make up a paltry 1.2\% of
all tracks in SCRG. We are interested in what tracks would produce
recommendations to AI tracks.
The recommendation behaviors of human-created and AI-generated
tracks are very different\footnote{Spotify's recommendation system is powered
  by a complex machine-learning model that accounts for many factors, such as
  similarity in audio characteristics, artists, and users' listening
  histories and habits. Therefore, we perform additional analysis to rule out
  similarity in music genre as the leading causein our observations, which we
  detail in Appendix~\ref{app:genre-factor}.}: 84\% of human-created tracks
do not recommend any AI tracks, while 93\% of AI tracks recommend
other AI tracks. Furthermore, each AI track has recommended 36 other AI
tracks on average, mapping to nearly 40\% of its total recommendations.
These observations remain when we remove recommendations between songs
released by the same artist.

Such a stark contrast between AI-generated and human-made music is not
entirely surprising, given that AI tracks often share similar
characteristics, particularly when they originate from a small number
of leading generators. More on the recommendation network
is in Appendix~\ref{app:simulating-ai-listening} and Figure~\ref{fig:ai-recommendation-network}.

\begin{figure}[t]
  \centering
  \includegraphics[width=0.9\linewidth]{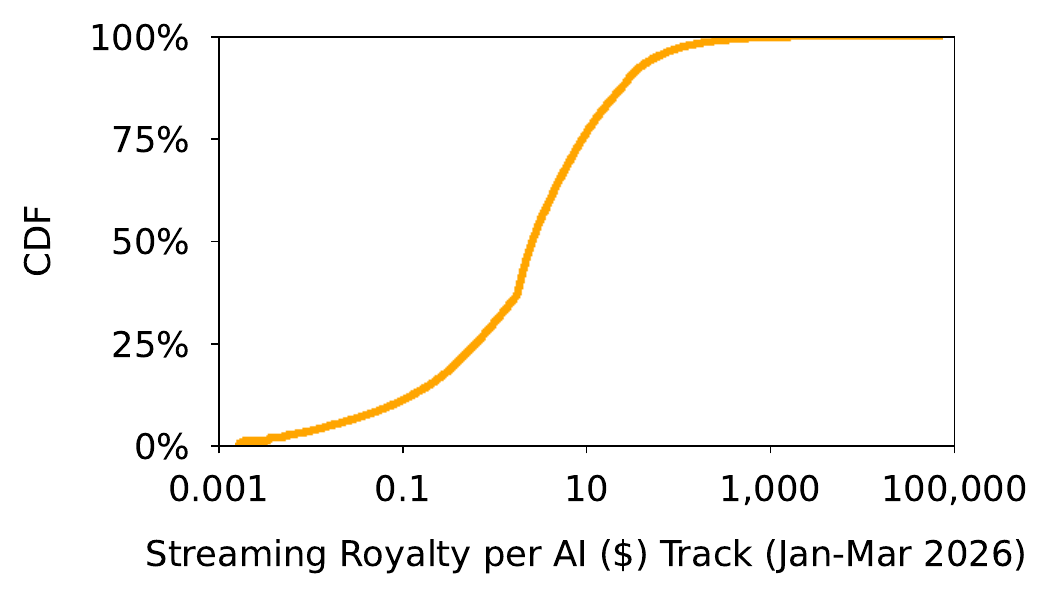}
  \caption{\swuedit{Only 7.15\% of the most popular AI tracks in our dataset
      generated enough play counts between Jan. and May 2026 to be
      monetized. We plot the CDF of estimated royalties among this small
      group of AI tracks. Even here, the majority of AI tracks earned less than \$10.}}
  \label{fig:ai-music-revenue}
\end{figure}

\subsection{Streaming Royalty and Profitability}
\label{subsec:ai-music-engagement-revenue}
Next, we study the economic outcomes of AI music (and slop) on streaming
platforms, by analyzing streaming royalties earned by AI tracks and whether
(and how) their royalty patterns could sustain a shadow industry similar to
the spam and fake pharma industries.

Today, it is extremely easy (and low cost) to generate and publish large
volumes of AI-generated tracks. Simple calculation show a maxed out Suno
membership would allow 6000 tracks per year for less than \$0.02 per track
(\S\ref{subsec:generators}), and all of them could be distributed for an
annual fee of \$26 or less (Table~\ref{tab:ai-music-and-distributors}). The
grand total would be \$126 per 6000 tracks, or two cents per track.

\para{Royalty earned by AI music.} Spotify does not make
royalty earnings of individual tracks public, and our datasets do not include such
metadata by default. However, based on Spotify's official
documentation~\cite{spotify-monthly-streamshare}, royalty payouts are derived from
play count via a monthly streamshare model. For each track, Spotify calculates the ratio of
the track's play count in a given month to the platform's total play
count for all tracks for the same month. Royalties are then distributed proportionally
from a total pool based on these ratios. For our analysis, we estimate royalties using
use the crawled playcounts for January to May 2026 from \S\ref{subsec:ai-slop-engagement}.

There are several limitations of this royalty estimation.  First, royalty
sharing is impacted by geographic location of listeners. Since this level of
detail is not available to us, we make the simplifying assumption that all play counts
are equal in our estimate. Second, Spotify only enables monetization for
tracks with at least 1,000 plays within the last 12 months. We lack precise
month to month data, so we include all tracks with more than 1,000 total
plays as part of our monetization estimate. Since the large bulk of AI tracks
were added less than a year before our study, we do not believe that this
decision will greatly impact our royalty estimate for AI songs.

Using this monthly play count data, we compute royalties earned by
each track. \swuedit{For each month from January to May, AI music contributed to 
$\approx 0.3$\% of the monthly royalty payout. Assuming the total royalty 
pool equals the average monthly royalty in 2025 reported by
Spotify~\cite{spotify-total-royalties}, this equates to roughly \$13.4
million USD total over this 5 month time period.}

\begin{figure}[t]
  \centering
  \includegraphics[width=0.95\linewidth]{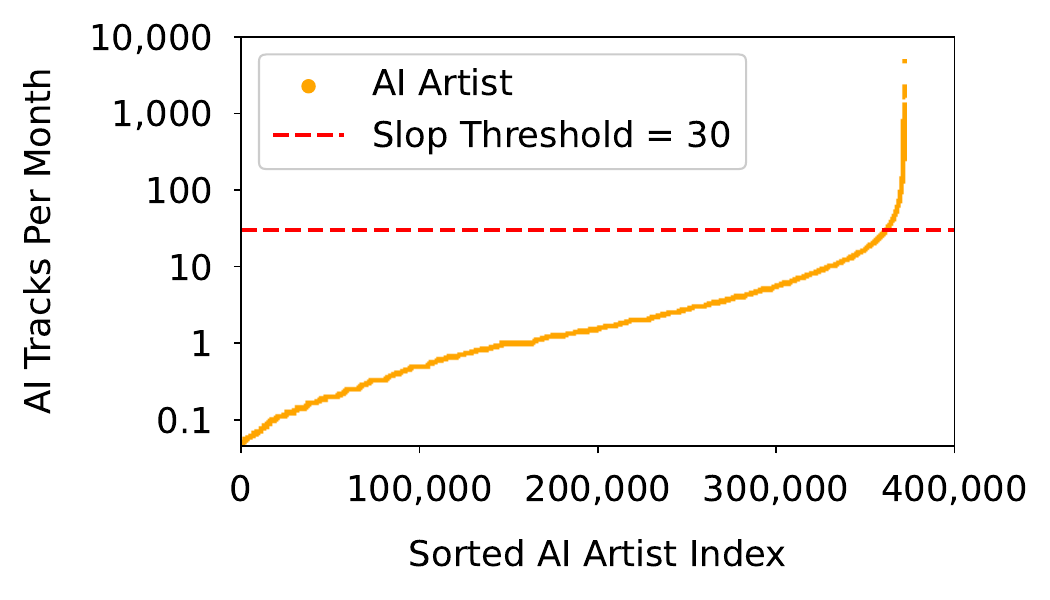}
  \caption{We plot the average number of AI tracks released for each
  AI musician. A majority of AI musicians exhibit benign upload behaviors,
  and we believe the small portion that upload $>$30 tracks per month
  are slop musicians.}
  \label{fig:ai-slop-threshold}
\end{figure}

\para{Royalty inequality.} Popularity and income in the streaming music
industry is known to be highly skewed, where a small number of extremely
popular songs capture the majority of revenue/streamshare. Given the
  ``spray and pray'' approach we observed with AI slop artists, we seek to
understand whether such inequality also exists among AI music.

Figure~\ref{fig:ai-music-revenue} plots the cumulative distribution of
royalties earned from January to May 2026 for the most popular AI tracks,
which we define as those with at least 1 non-negligible play (and thus earn
royalties). This subset only represents the top 7.15\% of all AI tracks, and
even within this subset, 76.5\% earned less than \$10. Only 1,632 AI tracks
(0.27\%) earned more than \$1,000.

\subsection{Separating AI-musicians from Slop}
\label{subsec:separating-slop}
While still small in
absolute terms, it is clear that AI-generated music is producing tracks that
are being monetized. Our next step is to examine this more closely, to answer
the key question: are AI artists that generate slop rewarded for their
efforts with hit songs that produce profits?

\begin{figure}[t]
  \centering
  \includegraphics[width=0.9\linewidth]{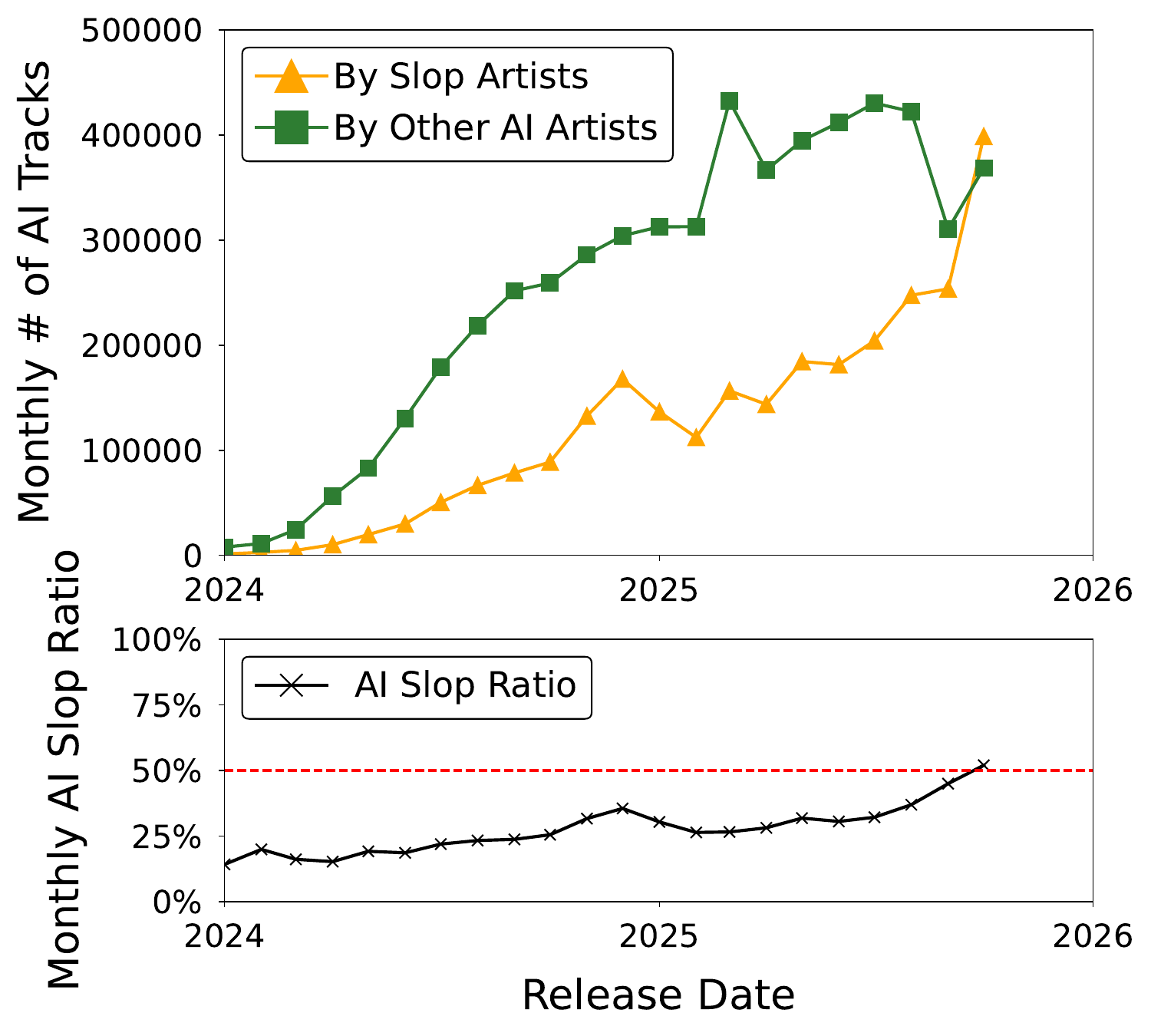}
  \caption{Growth of AI-generated music separated into ``slop'' artists and
    ``non-slop'' AI artists, shown as number of new AI-generated tracks
    arriving each month, and as a ratio of AI slop music to all AI music.}
  \label{fig:ai-slop-arrival}
\end{figure}

\para{Classifying AI music slop.}
To answer this question, we make an effort to distinguish between AI
musicians who generate large volumes of low quality slop, and AI musicians
who use generative AI to assist the creative process. Without a way to
empirically determine the ``quality'' of an AI-generated track, we use the
frequency of generation as our main criteria for identifying slop generators
(see \S\ref{subsec:ai-music-properties}). \swuedit{More specifically, we use 30 tracks
per month as the threshold for identifying AI-slop generators, AI musicians
focused on producing high-volumes of low-effort music. This threshold was
chosen conservatively based on the behavior distribution of AI artists (see Figure~\ref{fig:ai-slop-threshold}, only 2.7\% of AI artists meet this requirement.)}

\para{Growth of AI music slop.}
We classify all of our AI-musicians as either ``slop artists'' or ``other AI
artists'' using this threshold. We then plot over time the arrival of AI
tracks generated by either slop artists or other AI
artists. Figure~\ref{fig:ai-slop-arrival} shows a very interesting trend,
showing that for much of 2024 and early 2025, AI music generated by non-slop
artists significantly outpaced AI music produced by slop artists. However,
the amount of AI music slop rises quickly in the second half of 2025, and
surpasses legitimate AI musicians by October 2025. This is visible when we
look at the AI slop ratio over time, which shows AI music slop becoming the
majority of AI music by October 2025.

\begin{figure}[t]
  \centering
  \includegraphics[width=0.9\linewidth]{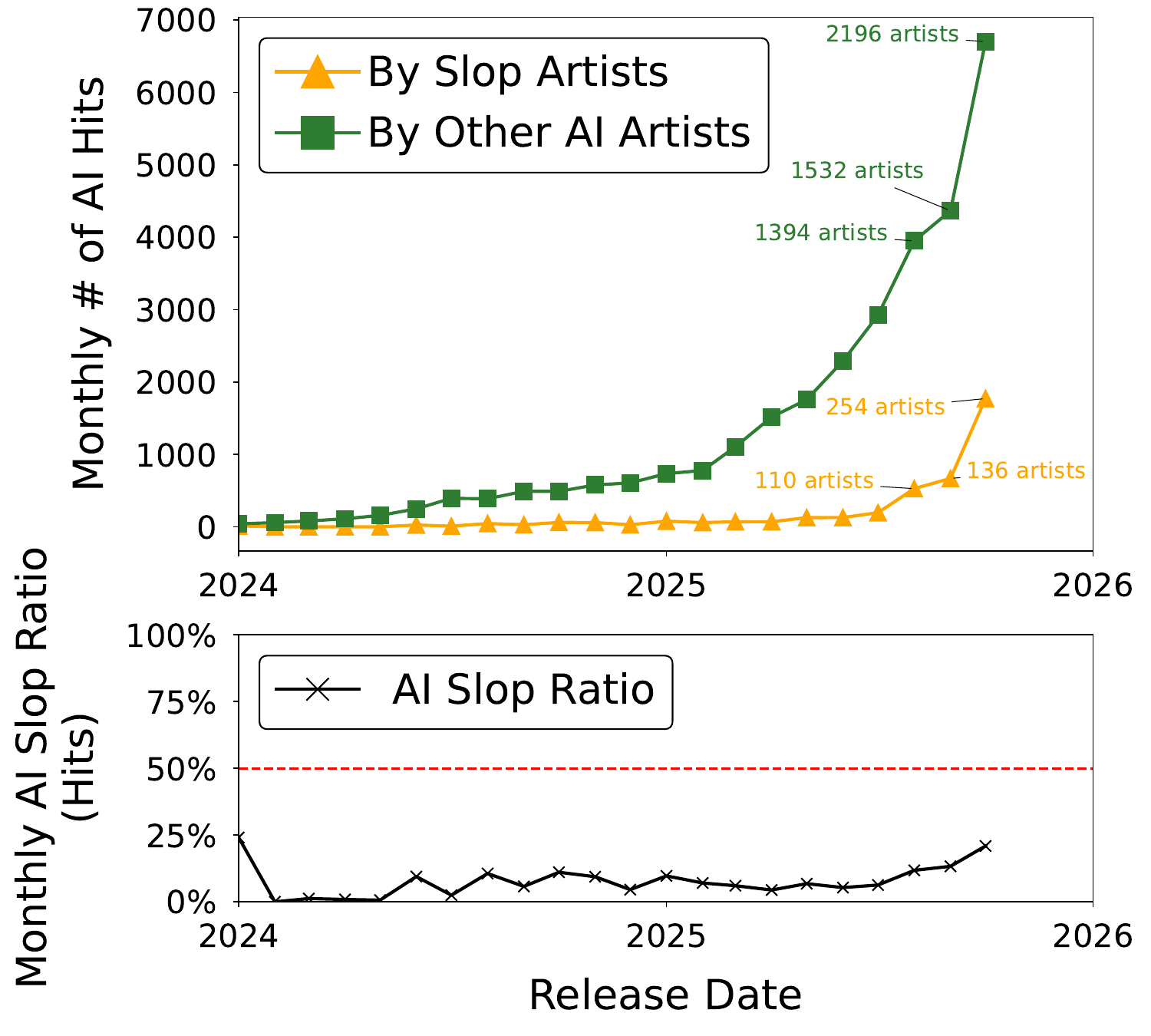}
  \caption{Growth of ``hits'' produced by AI-generated music, divided into
    ``slop'' and ``non-slop'' AI artists, sorted by month of initial release.}
  \label{fig:popular-ai-songs-arrival}
\end{figure}

Next we perform a similar analysis, but focus on profitable ``hit'' songs,
which we define to be AI tracks that produce at least 6,000 listens per
month.  By our estimates, this threshold generates sufficient monetization
(~\$200/year) to produce reasonable
profits. Figure~\ref{fig:popular-ai-songs-arrival} plots the month of arrival
for all AI tracks that become ``hits'' by our definition. Again, we plot
absolute count over time as well as the ratio of contributions from slop
artists.

Here, we observe that both AI artists and slop artists are accelerating in
their generation of hit tracks, but at dramatically different rates. Numbers
from the latest month show 2196 benign AI artists produced 6703 hits. In
contrast, 254 slop artists produced 1772 total hits, despite total number of
slop tracks outnumbering benign AI tracks.
While these numbers are paltry compared to
``hits'' generated by human musicians (monthly average of 72,000 in late
2025), it suggests that slop artists, even though they are less successful
than well-intentioned AI musicians, could able to generate enough profit to
sustain their operations.

\begin{figure}[t]
  \centering
  \includegraphics[width=0.95\linewidth]{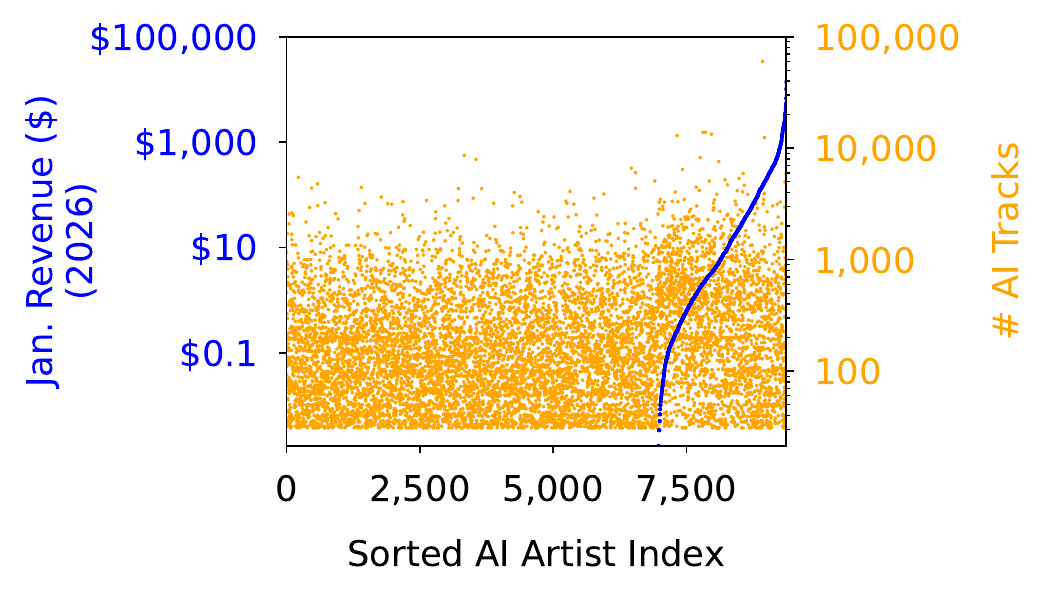}
  \caption{Distribution of revenue for slop artists, sorted by increasing
    revenue, alongside number of tracks released on secondary axis.}
  \label{fig:ai-slop-revenue-distribution}
\end{figure}

\para{AI slop profitability.} We now look closer into the population of slop
artists to understand how successful they are at generating and monetizing
slop. For this, we look at a single month period in January 2026, and analyze
the monthly revenue for all 9,891 AI artists who qualify under our threshold
for slop generator (those who release at least 30 tracks per month). We plot
in Figure~\ref{fig:ai-slop-revenue-distribution} revenue for each slop
artist, along with the number of tracks they released that month.

The large majority of slop artists fail to generate any revenue. But more
surprisingly we see that there is no strong correlation between tracks
released and top revenue generators. The top revenue generators in our slop
artist group cover a wide range of slop volume, including some who generate
millions of listens (and hundreds of dollars per track), despite only
releasing 30-50 tracks per month. Clearly some slop artists have figured out
how to generate significant revenue per track. Visual inspection of these
artists show no obvious pattern or common traits, other than that most are
international artists. We have no obvious explanation, except speculation
that some portion of these extremely popular artists could be artificially
inflating play counts using automated bots.

At the above rate in which AI slop music is generating revenue, current
streaming platforms appear to present a valid opportunity to earn significant
royalties by flooding streaming platforms in hopes pf generating a (rare) AI
hit that pays off in royalties. This is a similar incentive model that drives
and sustains large-scale spam campaigns~\cite{spamalytics}.

%% file: src/ai-artist.tex
\section{Distribution and Throttling of AI Slop}
\label{sec:distribution}

Our analysis of AI-generated music on Spotify shows that music slop is
rapidly growing on Spotify, and threatening to overwhelm both human musicians
and non-slop AI musicians. AI music slop indeed exhibit typical
characteristics associated with AI slop: low-cost, 
high-volume, low-engagement content. To prevent slop from overwhelming
streaming platforms, the industry must apply throttling to limit the growth
of music slop. 

This leads to the question, what kind of mechanisms in today's music
ecosystem could have a realistic throttling effect on AI music slop? To answer
this, we must first understand the pipeline of AI music, from generation to
distribution to streaming on user-facing platforms like Spotify.
We do so through {\em proactive} experiments where we track the flow  of new
AI tracks, from generation to distribution to streaming. By observing the
progression of these tracks through the music publishing pipeline, we gain
insights into the roles distributors, streaming platforms and their
policies can play in limiting AI music slop.

\subsection{Generating and Uploading AI Music}
\label{subsec:uploading-ai-music}
Next, we describe our experiences as we generate and publish our
own AI songs onto streaming platforms.

\begin{table}[t]
  \centering
    \resizebox{0.468\textwidth}{!}{
      \begin{tabular}{@{}ccccccc@{}}
      \toprule
      \textbf{Distributor} & \textbf{Cost} & \textbf{\begin{tabular}[c]{@{}c@{}}Royalty\\ Tax \%\end{tabular}} & \textbf{\begin{tabular}[c]{@{}c@{}}AI Music\\ Allowed\end{tabular}} & \textbf{\begin{tabular}[c]{@{}c@{}}Detected\\ AI\end{tabular}} & \textbf{\begin{tabular}[c]{@{}c@{}}\# AI Songs\\ Approved\end{tabular}} & \textbf{\begin{tabular}[c]{@{}c@{}}Review\\ Time\end{tabular}} \\ \midrule
      CD Baby & \$9.99 / track & 9\% & {\color{darkred} N} & {\color{darkred} 0/8} & {\color{darkgreen} 8/8} & 14 days \\
      Horus Music & \$23.03 / year & 0\% & {\color{darkred} N} & {\color{darkred} 0/8} & {\color{darkred} 1/8} & 3 days \\
      Offstep & \$18.00 / year & 0\% & {\color{darkred} N} & {\color{darkred} 0/8} & {\color{darkgreen} 8/8} & 2 days \\
      Soundrop & \$4.99 / track & 15\% & {\color{darkred} N} & {\color{darkred} 0/8} & {\color{darkgreen} 8/8} & 6 days \\
      TuneCore & \$22.99 / year & 0\% & {\color{darkred} N} & {\color{darkred} 0/8} & {\color{darkgreen} 8/8} & 5 days \\ \midrule
      Amuse & \$23.99 / year & 0\% & {\color{darkgreen} Y} & {\color{darkgreen} 4/8} & {\color{darkgreen} 8/8} & 2 days \\
      DistroKid & \$24.99 / year & 0\% & {\color{darkgreen} Y} & {\color{darkred} 0/8} & {\color{darkgreen} 8/8} & 1 day \\
      Ditto & \$19.00 / year & 0\% & {\color{darkgreen} Y} & {\color{darkred} 0/8} & {\color{darkgreen} 8/8} & 1 day \\
      Landr & \$23.99 / year & 0\% & {\color{darkgreen} Y} & {\color{darkgreen} 6/8} & {\color{darkgreen} 8/8} & 5 days \\
      Routenote & free & 15\% & {\color{darkgreen} Y} & {\color{darkred} 0/8} & {\color{darkred} 0/8} & 1 month \\
      UnitedMasters & \$19.99 / year & 0\% & {\color{darkgreen} Y} & {\color{darkred} 0/8} & {\color{darkgreen} 8/8} & 5 days \\ \bottomrule
      \end{tabular}
    }
  \caption{Our findings after uploading AI songs to 11 music
    distributors. Notably, even distributors that prohibit AI music (CD Baby,
    OFFstep, Soundrop, and TuneCore) approved AI songs and sent them to
    streaming platforms.} 
  \label{tab:ai-music-and-distributors}
\end{table}

\para{Generating AI tracks.} We create a number of distinct AI artist
personas, each with their own small library of 8 fully generated AI songs,
generated by 4 different generators. As mentioned in
Section~\ref{subsec:generators}, we used the two most popular commercial
generators (Suno and Udio), and the two highest-quality open-source
generators (DiffRhythm and ACE-Step). For each generator, we generated one
instrumental and one lyrical song, using generic AI generated text input
prompts, with lyrics generated by LLaMA 3.1. Further details on the
generation process can be found in Appendix~\ref{app:ai-track-generation}.

We submitted each generated track as an album single (album with one track),
and used FLUX~\cite{flux2024} to generate an album art cover. To meet the
resolution requirements for distributors, we used SDXL's 4x image
upscaler~\cite{sdxl}. Our decisions are designed to pass automated checks
against sybils/spams while minimizing effort.  We only uploaded 8 AI songs
per distributor to minimize any negative impact of our songs on human
musicians.

\para{Selecting music distributors.} To distribute music to streaming
platforms, we need to choose a set of distributors to work with.  From the
list of 18 distributors identified by indie
musicians~\cite{different-distributors}, we exclude those that required ID
verification or exclusive partnerships, leaving 11 distributors to work
with. This list includes Ditto and TuneCore, distributors that have worked
with well-known independent artists like Chance the Rapper and Russ.  For
each distributor, we identify their policy regarding AI-generated music
(explicitly allow or prohibit). We show this information in
Table~\ref{tab:ai-music-and-distributors}\footnote{Recently, CD Baby and
  Horus have changed their policy to allow AI music.}, along with cost for
distribution and royalty sharing agreements.

We are aware that music distributors often share data about song metadata to
avoid duplication or copyright infringement. Streaming platforms like Spotify
and distributors like DistroKid reportedly employ audio fingerprinting to
ensure the same songs are not duplicated on their
platform~\cite{spotify-fingerprinting}. Of course, rejections based on
overlapping audio fingerprints would disrupt our experiment. For each music
distributor, we assign one of our unique artist personas, and upload a {\em
  unique} subset of 8 songs to each distributor.

\para{Uploading and observing.}  For all 11 music distributors, we manually
upload every track, with a maximum of 2 per day per distributor (to avoid
triggering high-frequency rate limits). We selected 6 different music
streaming platforms as destinations for our music: Spotify, Apple Music,
Deezer, Amazon Music, YouTube Music, and Tidal. These represent a majority of
the most popular music streaming apps~\cite{streaming-industry}. We performed
these experiments between November and December 2025, and manually verify in
January 2026 to determine if tracks successfully made it to each platform.  

\subsection{AI Policies and Enforcement}
\label{subsec:ai-music-policies}
We first discuss distributors' and streaming platforms'
AI music policies, and their possible enforcement.

\para{Distributors disagree on AI policies.}  Uploaded music must be approved
by distributors before they can be sent to streaming platforms. As we see in
Table~\ref{tab:ai-music-and-distributors}, current music distributors are
split on whether users should be allowed to upload fully AI-generated
songs. All 11 distributors we evaluated mentioned AI usage in their user
policies, but varied significantly in what they mention. Some disallowed any
use of AI\footnote{\swuedit{Note that no distributors we tested prohibit 
uploading AI-generated music in their terms of service. Rather, each distributor maintains
its own policy on whether they distribute AI music to streaming services. Our study 
evaluates each distributor's success implementing their stated policy.}}, 
while others disallowed only fully generated AI songs. All
distributors require uploaders to fully own the rights to the songs they
upload, but there is no consistent definition of how that applies to AI-generated 
songs. The only consistent AI policy we found is prohibiting the
usage of AI to impersonate existing musicians.

\para{Streaming platforms lack stance on AI.}  Once music is approved by
distributors, it still must be approved by the streaming platforms
themselves. None of the 6 major platforms we tested (Spotify, Deezer, Apple
Music, Amazon Music, YouTube Music, Tidal) mention whether fully AI-generated
songs are allowed on their platform. However, Spotify and Deezer have taken
steps to address AI music. Deezer now labels AI music and demonetizes
fraudulent AI content, while Spotify claims to have removed millions of
``spammy'' tracks. Apple Music's policy recently added AI transparency tags, but
are completely dependent on distributors' ability and willingness to detect
and label AI music~\cite{applemusic}.

\para{Enforcement.}  We found mixed evidence on enforcement of
these AI policies. None of the 4 distributors who
prohibit AI submissions explain how they enforce it. In fact, in our
set of 11 distributors, we only found evidence that Amuse and Ditto use an AI
music detector, but both allow AI music.

\begin{figure}[t]
  \centering
  \includegraphics[width=0.9\linewidth]{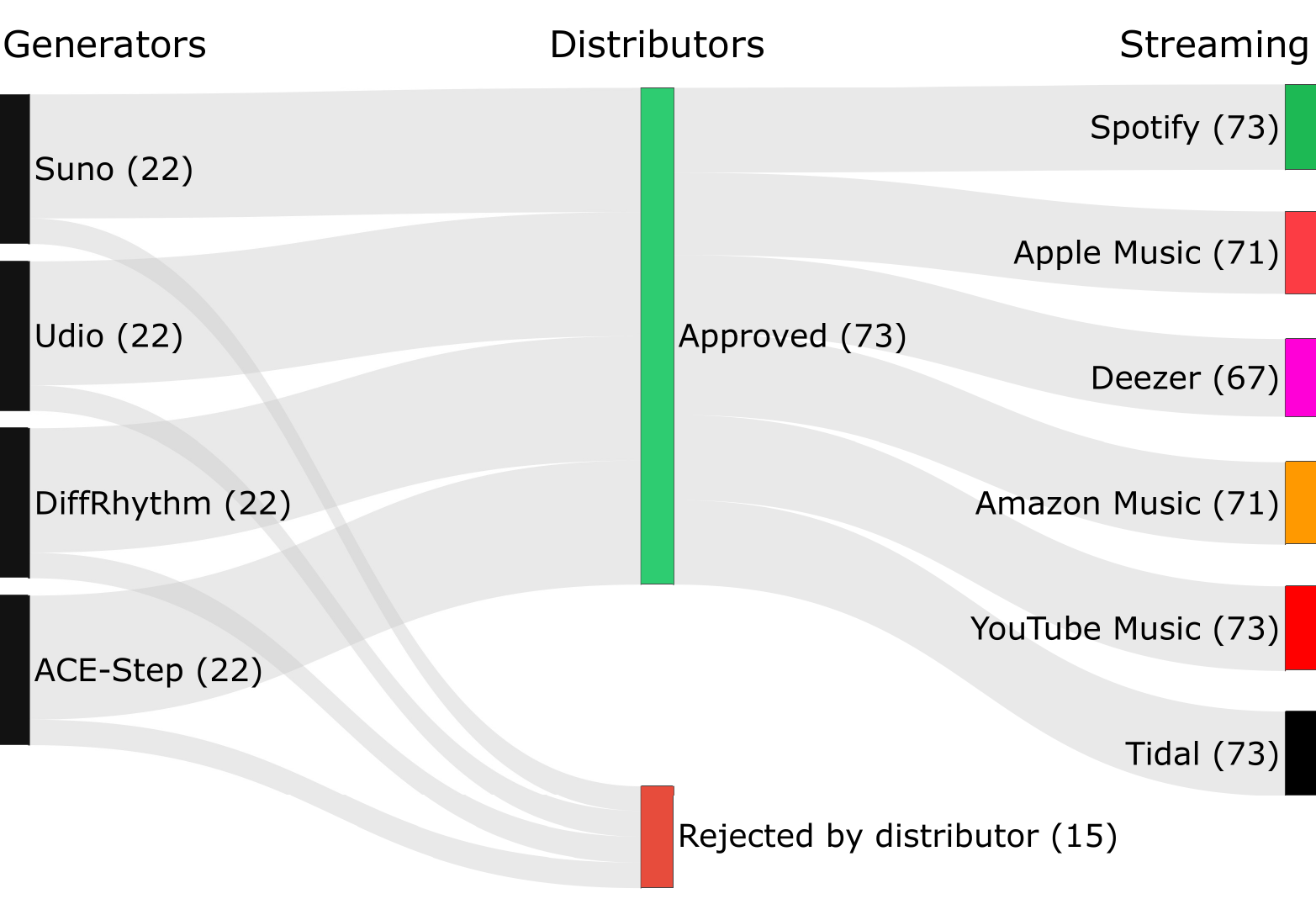}
  \caption{Results of our end-to-end AI music uploading. Rejections mainly occurred at the music distributor. Once songs get approved by distributors, they are rarely rejected by streaming platforms.}
  \label{fig:ai-music-publishing-pipeline}
\end{figure}

\subsection{Results of Our AI Music Uploads}
\label{subsec:aiuploadresults}

\para{Distributor approval.}  We list the results of our uploaded AI music
tracks in Table~\ref{tab:ai-music-and-distributors}. Out of a total of 88
tracks uploaded, a total of 15 songs were rejected by 2 distributors (Horus
and Routenote). Rejected songs did not correlate with the AI
generators used.  Either all are approved by the target distributor, or
almost all are rejected. Notably, 4 out of the 5 distributors that
explicitly prohibited AI music uploads approved all AI songs we
uploaded. When rejecting our uploads, Horus Music and RouteNote did so under
a blanket claim that our songs did not meet their platform's
requirements. Neither marked the rejected songs as AI-generated.

\para{Lack of reliable AI detection.}  Based on their documentation, all
distributors but one (DistroKid), employ a review process that involves
review by a moderation team. DistroKid uses a fully automated review
pipeline. Two distributors that prohibit AI music (CD Baby and OffStep),
explicitly specify that they use humans to review every submission. Still,
none of them labeled our uploads as AI generated. Only two distributors
(Amuse and Landr) identified a subset of our uploads as AI-generated. Recall
from \S~\ref{subsec:distribution} that Amuse and Ditto are known to work
with SubmitHub, a commercial AI music detector. Yet Amuse detected only 4 out
of 8 tracks as AI, and Ditto detected none.  Detection clearly is a
significant hurdle to managing AI music, and we study it in detail next in
Section~\ref{sec:ai-music-detection}.

\para{Streaming platforms rarely rejects tracks approved by distributors.}
In Figure~\ref{fig:ai-music-publishing-pipeline}, we show how many of our AI
tracks appear on streaming platforms after receiving approval by
distributors.  Spotify, YouTube Music, and Tidal accepted all songs approved
by distributors, while Apple Music and YouTube Music each rejected 2 songs
(out of 73). While we received no detailed reasons for these rejections, we
infer it was because these songs did not pass Apple/YouTube's
platform-specific quality assurance tests.

Although it looks at first glance like Deezer rejected 6 of our tracks, a
closer look revealed that Landr was responsible. Landr incorrectly assumed
that Deezer does not accept AI-generated music, and thus did not send them
the 6 tracks it identified as AI-generated, despite approving them for
distribution to other platforms.

\para{Possible throttle points and challenges.}  Our experiences creating and
uploading AI music reveal a number of insights into life cycle of
AI-generated music and potential throttle points for AI music slop. We
consider each potential throttle point in turn:
\begin{packed_itemize}
\item {\em Music Generators}. AI music generators profit from AI music creation,
  and are unlikely to limit the usage of their model. However, proprietary
  generators like Suno could potentially raise their prices, increasing the
  base cost per AI song and limiting potential profit from music slop.
\item {\em Distributors}. Distributors have the ability to curate and filter songs
  during distribution based on policies. They are in the best position to
  properly label AI music, assuming they have a reliable detection
  mechanism.
\item {\em Major Labels/Distributors.} Major record labels like Warner Music
  distribute music from their own signed artists. Major labels have more
  resources and are more likely to employ strong detectors and filters to
  apply their own policies on AI generated music~\cite{slopmusicumg}.
\item {\em Streaming Platforms.} Platforms like Spotify and Apple Music are
  currently evolving their policies towards AI-generated music towards
  transparency via labeling. Given the monetary
  incentives~\cite{slopmusicmimicry}, they are more likely to allow users to
  choose to listen or avoid AI music than they are to apply proactive
  filtering.
\item {\em Users.} Personal opinions on AI generated music (including slop
  and non-slop) are likely to vary significantly across users and evolve over
  time. Given a choice, we expect many users will opt to filter or moderate
  their consumption of AI generated music~\cite{slopmusicusers}.
\end{packed_itemize}

While we expect the issue of moderating music slop to be a complex and
potentially contentious issue moving forward, one thing is clear: any hope of
effective moderation of AI music slop requires reliable and accurate
detection of music slop (and AI-generated music in general). We study this
challenge in detail next.

%% file: src/detection.tex
\section{Detecting AI-Generated Music}
\label{sec:ai-music-detection}

One key insight from our experiments with AI-generated music
(\S\ref{sec:distribution}) is that reliable detection of AI-generated music
is critical to providing transparent choice to users and potentially
throttling music slop. Without reliable detection and labeling, any policies
regarding AI-generated music by distributors, streamers, or users will not be
enforceable. 

Here we study this problem in detail, and evaluate automated 
detection of AI-generated music.
We begin by evaluating the detection efficacy (under both benign and
adversarial conditions) of four open-source detection methods, including one
that is the basis for Deezer's commercial detector (see
\S\ref{subsec:generators}). We also use end-to-end tests to evaluate the
robustness of the commercial Deezer detector.

Our results show that AI music detection is still not a fully solved problem. While
one approach (Deezer's Fourier detector) shows high accuracy under benign
conditions, it is easily bypassed by applying simple preprocessing operations
to the music file, including lossy compression, pitch shift, reverb, and
AI-reconstruction. Adversarial training restores robustness, but not against
all possible attacks.

\begin{figure}[t]
  \centering
  \includegraphics[width=0.85\linewidth]{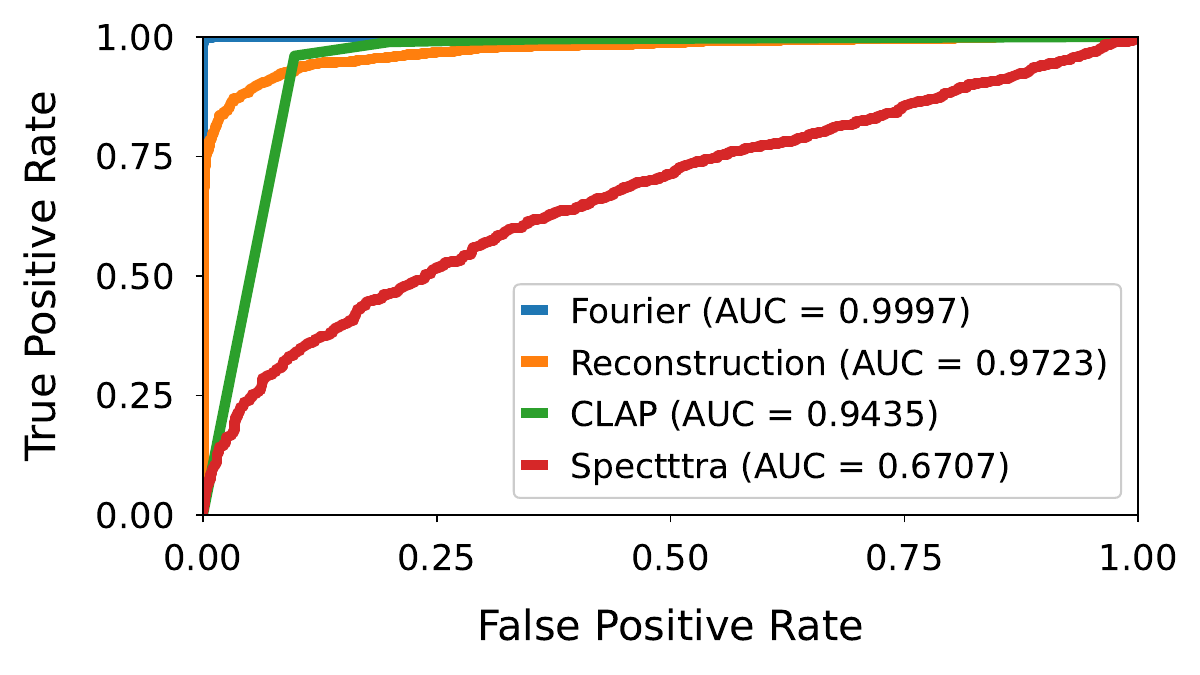}
  \caption{Performance of AI music detectors shown as ROC curves. The
    Fourier-based detector (used by Deezer and Quicksilver) is
    nearly perfect at distinguishing AI songs generated by Suno/Udio from human-created songs.}
  \label{fig:detector-roc-curves}
\end{figure}

\subsection{Automated Detection}
\label{subsec:ml-detection}
We first examine the accuracy of ML-based AI music detectors and consider
factors that are most likely to affect robustness. Our experiments use test
data from multiple music generators, as well as transformed samples 
that serve as adversarial attacks.

\para{Dataset and methodology.} For our test set, we randomly sample and
download a total of 1,000 songs generated by two most widely used music
generators, Suno~\cite{suno-dataset} and Udio~\cite{udio-dataset}. They are
the only commercial models\footnote{Detection of open-source generators can
  be found in Appendix~\ref{app:diffrhythm-acestep}.} producing full-length
songs. For human-made music, we randomly sample 1,000 songs released before
2021 and download 30 second previews for testing.

 We evaluate four SOTA ML-based detectors: 
\begin{packed_itemize}
\item \textbf{Spectttra~\cite{sonics}:} transformer trained on
  spectrograms to detect Suno/Udio songs.
  
\item \textbf{CLAP-based~\cite{ai-arms-race}:} hierarchical
  classifier trained on CLAP embeddings to detect Suno/Udio songs.
  
\item \textbf{Reconstruction-based~\cite{cnn-reconstruction}:} CNN
  trained on spectrograms of reconstructed audio to detect VAE
  usage in AI music generation.
  
\item \textbf{Fourier-based~\cite{fourier-peaks}:} Linear classifier trained
  to identify synthetic peaks in spectrograms of Suno/Udio songs.  Deezer's
  built-in AI labeler~\cite{fourier-peaks} and Quicksilver~\cite{quicksilver}
  are based on this architecture. For our tests, we use an open source
  version of \cite{fourier-peaks}.
\end{packed_itemize}

For each detector, we use code/models released by the authors, and if
unavailable, we reuse their code release to train a model using separate
training data from Suno and Udio (with no overlap with our test set).  For
each detector, we convert the audio to the same format used to train the
detector before evaluation.

\para{Detection performance.}  We evaluate each detector on the test set
(1,000 AI, 1,000 human) and plot performance via ROC curve in
Figure~\ref{fig:detector-roc-curves}. Overall, the Fourier-based detector
outperforms the other three, with 99.6\% accuracy. It also maintains an
extremely low false positive regime, i.e. 99.6\% TPR at 0.1\% FPR. We also
tested Quicksilver~\cite{quicksilver}. As another implementation of the
Fourier-based approach, it showed (unsurprisingly) very similar performance
(99.75\%) accuracy. Reconstruction-based detector is the next best, achieving
68\% TPR at 0.1\% FPR.

ML-based detectors are known to be vulnerable to adversarial attacks, which
authors of both the Fourier and CLAP-based detectors have confirmed in their
work~\cite{fourier-peaks,ai-arms-race}. We verified using experiments
that lossy compression can bypass both detectors.  Next, we show that similar
audio processing techniques can also be used to bypass Deezer's commercial AI
music detector, which is currently being used to label its tracks.

\begin{table}[t]
  \centering
    \resizebox{0.35\textwidth}{!}{
        \begin{tabular}{@{}cc@{}}
          \toprule
          \textbf{\begin{tabular}[c]{@{}c@{}}Evasion 
                    Technique\end{tabular}} &
                                              \textbf{\begin{tabular}[c]{@{}c@{}}Success
                                                        Rate\end{tabular}} \\ \midrule 
          MP3 Compression& 5/5 \\
          Pitch Shift & 5/5 \\
          Reverb & 5/5 \\
          Reconstruction & 5/5 \\ \bottomrule
        \end{tabular}
    } 
    \caption{All 5 AI-generated songs modified by each of the 4 techniques
      successfully bypassed Deezer's AI detector.}
  \label{tab:evading-deezer-detection}
\end{table}

\para{Evading Deezer.}  While we have no direct access
to Deezer's AI detector, we test it by utilizing the distribution pipeline
established in \S\ref{sec:distribution} to upload music to Deezer and
manually observe their labels. Specifically, we generate 20 unique AI songs
using Suno (the best performing generator), and modify each song by selecting
one of 4 audio processing techniques: lossy compression, pitch shift, reverb,
and audio reconstruction using an autoencoder.  For each of the 4 evasion
techniques, we upload 5 altered songs. Each song is then uploaded through
DistroKid (fastest review turnaround of 1 day), specifying Deezer as the
streaming target. Other than pitch shift, none of the evasion techniques
alter the audio in a distinguishable manner to human listeners.
Table~\ref{tab:evading-deezer-detection} shows that all 4 techniques
bypass Deezer's detector with 100\% success, which we suspect is likely because
Deezer's detector is trained without adversarial training methods.

\para{Adversarial training.} The open source implementation of the Fourier
detector shows similar vulnerability to evasion attacks. However, we are able
to apply adversarial training to further train the classifier
against each evasion method. Applying this to our local implementation
dramatically boosts robustness to evasion, restoring accuracy to 95\% against
all evasion attacks, with only a minor drop in benign classification
performance (dropping from 99\% to 98\%).

While this is a positive result, we note that adversarial training does not
generalize across different attacks. A robust detector adversarially
trained against a variety of evasion attacks could still be vulnerable to
unknown attacks.

\para{Disclosure to Deezer.}  We have disclosed this result to Deezer's research
team, who responded that they are currently working to improve
the robustness of their detector implementation.

%% file: src/economics.tex
\subsection{Economic Curbs for Slop Detection}
\label{sec:ai-music-economics}

Maintaining robust and accurate detection will be challenging over time. 
As music generators mature, we expect slop generators to find new ways to
evade detection.

Here, we outline a complementary approach that leverages the fundamental
motivation behind AI slop: money. Ultimately, AI music slop is about high
volume, low-cost content that floods the consumers, hoping that a small
subset of generated songs will become hits and generate revenue to fund the
entire campaign. The economics behind this is identical to that of
large-scale spam campaigns~\cite{spamalytics} and pharmaceutical
scams~\cite{pharmaspam}.  What makes this scenario different is that we have
stakeholders who are motivated to curb AI music slop, and who hold key
controls over the economics of AI music: {\em music distributors}.

\para{Initial observations.}
\label{subsec:ai-music-observations}
The current cost structure of music distribution is optimized for human
musicians (low volume, high quality), not potential AI slop producers (high
volume, low quality).  Most distributors charge a monthly/annual fee to
upload music (Table~\ref{tab:ai-music-and-distributors}). Properly amortized
over a year, third parties can reduce per track costs significantly below
\$0.02 (\S\ref{subsec:aiuploadresults}). This low cost structure designed to
support indie musicians to produce more tracks also enables low cost AI slop
production.

A key factor working in our favor is the relatively small number of music
distributors with relationships to streaming services. Any transparent
changes in costs will be clearly visible to streaming services. As the AI
music slop problem grows in severity, distributors can be incentivized to
change their cost structure and be perceived as preferred sources of higher
quality music.

\para{Shaping AI music using cost incentives.}
\label{subsec:shaping-upload-behavior}
There are multiple ways to curtail AI music slop. While we lack visibility
into the revenue and cost models inside music distributors, we can suggest a
number of potential changes to curb the volume of AI slop, each with clear
tradeoffs that might be more attractive to distributors than others.

\begin{packed_itemize}
\item Charge distribution costs per track, increase costs for those
  uploading large number of tracks while keeping costs stable for others. 
\item Increase distribution costs as a whole, decreasing probability of AI music slop turning a profit.
\item Increase per track costs as function of upload frequency, a
  disproportionally higher cost increase for those who upload large volumes
  of music.
\item Cap number of tracks per musician per time, forcing slop producers to
  create numerous fake accounts to scale.
\item Charge distribution costs per album, incentivizing AI slop producers to
  create large albums with hundreds of tracks, making tracks per album a
  reliable signal for detecting AI music slop.
\end{packed_itemize}

\para{Alternatives to assist AI music identification.} Finally, we note that
curbing AI slop does not help the accuracy of detecting AI music, a feature
that users seem to demand. One simple \swuedit{alternative} is to
consistently verify musician identity before distributing their music.  While
this is imperfect and can be gamed by bad actors, it nonetheless raises the
friction of uploading slop, and would strongly discourage organized spammers
from participation. Today, only 3 out of the 18 total distributors
from~\cite{different-distributors} require identity verification. Many AI
slop producers on Reddit comment that they avoid those distributors out of
fear of copyright violation claims or legal accountability.

%% file: src/conclusion.tex
\section{Conclusion}
\label{sec:discussion}
Our study leverages a large-scale measurement analysis on Spotify, end-to-end
experiments on music distribution, and analysis on existing detectors, to shed
light on the growth of AI music and music slop in today's music ecosystem.
We show AI music is easily injected into streaming platforms, and music slop
is dominating music created by both human musicians and benign AI
musicians. We identify robust detection as the key challenge to moderating
slop, and evaluate detection mechanisms. Deezer's Fourier-based approach is
accurate, but also vulnerable to adversarial evasion. We are able to improve its
robustness via adversarial training, and also suggest complementary economic
policies to improve detection accuracy.

\para{Limitations.}  We acknowledge that our analysis in
\S\ref{sec:crawl-eval} hinges on the accuracy of Deezer's AI music
detector. When we disclosed our results to Deezer, they told us they found little
evidence of users attempting to fool the detector. Regardless, our analysis
focuses on the overall trends of AI music, which we believe would
remain largely unchanged even if other strong detectors (like commercial
ones) were used instead.

%% file: src/appendix.tex
\section{}
\label{sec:appendix}

\input{src/ethics-open-science}

\subsection{Detailed Spotify Crawl}
\label{app:spotify-webcrawl}

\para{Crawling the largest connected component.}
To find the most important connected tracks, we leverage a BFS/snowballing methodology using Spotify's recommendations API to find new tracks. First, we started with a set of ``seed'' tracks, which we should be both comprehensive and diverse. Then, we repeatedly query Spotify recommendations API to establish connections (edges) from each seed track to new seed tracks. This process is repeated until there are no more new tracks recommended. Note that at each sampling layer, we additionally seeded the crawler with other songs released by the musician of the current track to expand our reach.

We seeded our crawl with 200,000 tracks spanning 75 different countries obtained through Spotify's official playlists, which represented the most relevant songs up to that date. Our crawl took 2 weeks to complete in November 2025, and returned $\approx$65 million unique track IDs. Then, using graph analysis, we computed the largest strongly connected component using the digraph constructed with each track as a node, and recommendations as directed edges. We find that there is a single large component (33 million tracks), and many tiny components, each with less than 20 tracks each. Based on this data, our SCSR dataset consists of two parts: 1) a database of 33 million tracks, where 6.8 million were released on/after 2024, and 2) a recommendation network of 33 million tracks (nodes) and 3.5 billion recommendations (directed edges).

\subsection{Detailed Deezer Matching Algorithm}
\label{app:deezer-match-algo}

As we described in \S\ref{subsec:datasets}, we labeled AI tracks in both our Spotify datasets by matching them to the Deezer platform, which already labels AI music. To accomplish this, we need to 1) match Spotify tracks to a corresponding one on Deezer, and 2) determine if the track on Deezer is labeled as AI. We were able to automate this process using two separate APIs.

\para{Searching and matching tracks on Deezer.}
For each Spotify track, we first query Deezer's official search API using combinations of the track's title, album title, and artist name to obtain a list of candidate matches. Then, we identify the correct candidate based on whether or not the two track's metadata match. Note that metadata between Spotify and Deezer is often not exactly the same (i.e., released at different times, title formatting, ..., etc.). As a result, we manually defined and tested a set of heuristics for a ``match.'' These rules are meant to be slightly fuzzy so as to catch any inconsistencies in metadata, which we found identifies a match roughly 80\% of the time.

\begin{packed_itemize}
    \item album title, track title, release year, and at least one artist name matches
    \item album title, track title, and all artist names match
    \item track title, release year, and all artist names match
    \item album title, track title, release year, and track duration match
    \item album title, release year, track duration, and copyright labels match
\end{packed_itemize}

Before we employed this matching algorithm on our entire dataset, we verified our approach by validating matches for 1,000 Spotify albums randomly sampled from SCRG. We manually verified that the correct Deezer album was found 99.6\% of the time. All incorrect matches were due to extraordinary circumstances (two different tracks/albums exist with the same name, duration, release year, and/or artist names).

\para{Labeling AI on Deezer.}
Once we established a match for a Spotify track on Deezer, our final step is to obtain the AI label that Deezer assigned to that matched track. Since Deezer labels AI at the album level, we just needed to identify the AI label for the album of the matched Deezer track. This labeling metadata can be publicly viewed on Deezer's web page for any album.

\subsection{Generating AI Tracks}
\label{app:ai-track-generation}
We generated AI tracks using as little human-input as possible. All content, including lyrics, was generated using a generative AI model. Below, we detail how we created each track, split by whether it was done using a closed/open-source model. For all tracks, cover art was generated via FLUX~\cite{flux2024} using prompts from ChatGPT, followed by SDXL's 4x upscaler~\cite{sdxl} to increase the resolution of the image.

\para{Suno/Udio tracks.}
Both Suno and Udio came with built-in capabilities to random generate song titles and lyrics. We used the randomize feature in both tools to generate full-length instrumentals and lyrical songs, and reused the song titles that were provided. 

\para{ACE-Step/DiffRhythm tracks.}
For the open-source models, we generated lyrics using LLaMA-3.1~\cite{llama} paired with song titles generated by ChatGPT. The prompt for the song was randomly chosen from a bank of 10 different song prompts also generated using ChatGPT.

\subsection{Genre as a Potential Factor for AI Music Connectivity}
\label{app:genre-factor}
In \S\ref{subsec:ai-music-accessibility}, we found that AI music is significantly more likely to recommend other AI music compared to human-made music. However, recommendation algorithms are complex, and there are many potential factors for this observation. One of the more intuitive reasons for why AI might recommend more AI is that AI tracks may share similar genres, which recommendation systems are likely to bias connections towards. Here, we conduct an experiment to rule out genres similarities as the primary reason for our AI music connectivity result.

To test this, we trained a logistic regressor on a sample of all edges (90/10 train-test split), equally divided between AI and human-created tracks from our recommendations network via the SCRG dataset, where inputs to the model $X_a$ is 1 if the source node is AI, 0 otherwise, $X_{gs}$ is the genre similarity of the source and recommended node, and $Y_a$ represents if the target node is AI.
\begin{equation*}
  Y_a = c_0 + c_1 \cdot X_a + c_2 \cdot X_{gs}
\end{equation*}
If genre similarity were to be the more important factor as to why AI tracks recommend other AI tracks, then $c_1$ should be much larger than $c_2$ after training. In Table~\ref{tab:logistic-regression-results}, we summarize the regression results. Our regressor achieves 81.39\% accuracy, and $c_1$, which is whether the source node is AI, has by far the highest positive coefficient ($p < 0.001$). At the very least, this confirms that the source track being AI has a much stronger effect on the recommended track being AI compared to genre similarity. 
\begin{table}[t]
  \centering
  \resizebox{0.35\textwidth}{!}{
    \begin{tabular}{@{}lccc@{}}
      \toprule
      \textit{\textbf{Predictor}} & Intercept & $X_a$ & $X_{gs}$ \\ \midrule
      \textit{\textbf{Coefficient}} & -5.173 & 4.555 & 0.623 \\ \bottomrule
    \end{tabular}
  }
  \caption{Logistic regression results for predicting whether a recommended track is AI-generated ($Y_a$). The dominant predictor (higher coefficient) is whether the source track is AI-generated ($X_a$), even accounting for genre similarity ($X_{gs}$). All predictors are statistically significant ($p < 0.001$).}
  \label{tab:logistic-regression-results}
\end{table}

\subsection{AI music and recommendations}
\label{app:simulating-ai-listening}
\begin{figure}[t]
  \centering
  \includegraphics[width=0.95\linewidth]{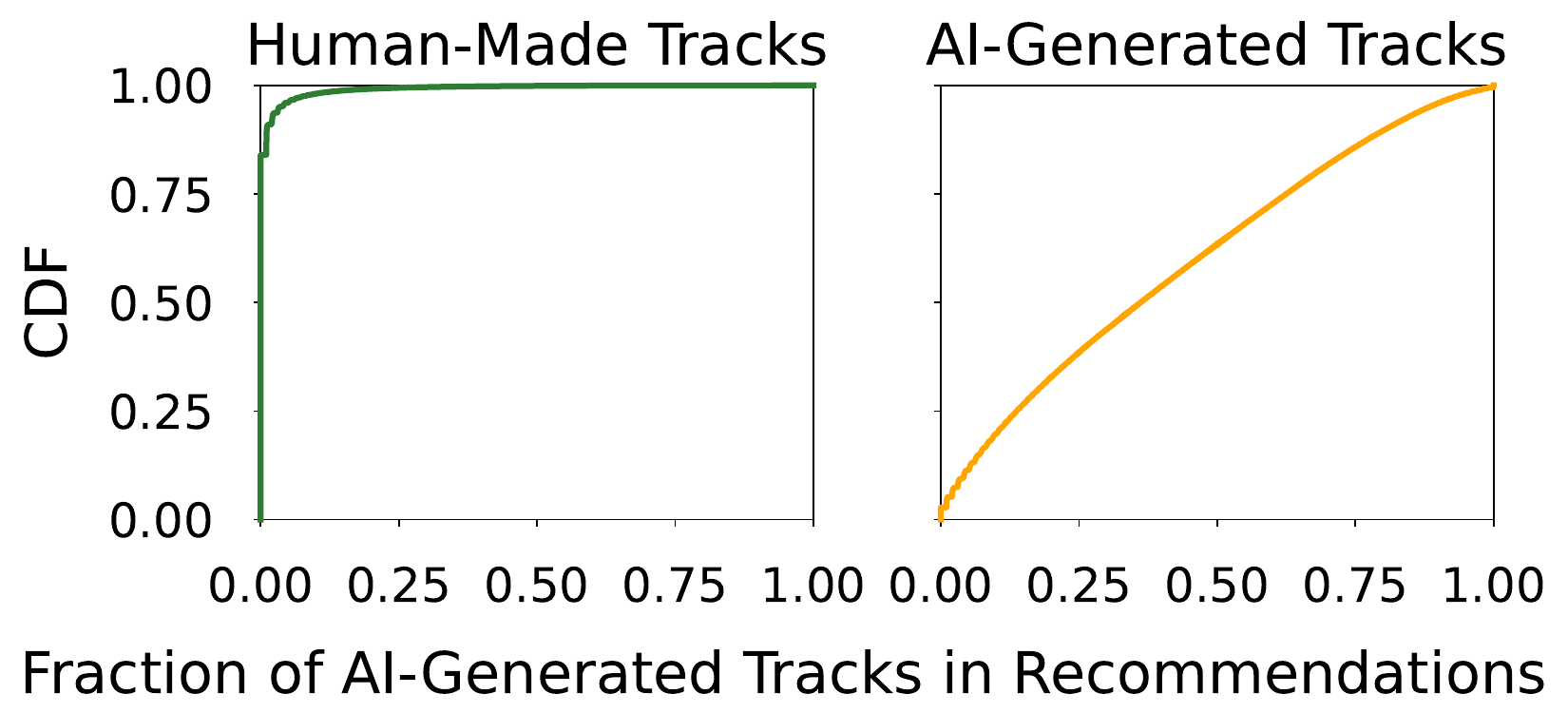}
  \caption{In Spotify's strongly connected  recommendation graph, 84\% of human-made
tracks never recommend any AI tracks;  97\% of AI tracks 
recommended other AI tracks.}
  \label{fig:ai-recommendation-network}
\end{figure}
Figure~\ref{fig:ai-recommendation-network}
plots the cumulative distribution of the fraction of AI tracks being
recommended by each track in the SCRG graph, where we report separately for
human-created tracks (left) and AI-generated tracks (right).

Because Spotify's recommendations encourage users to discover
  previously unheard tracks, we want to understand whether this system
  may eventually direct users to AI-generated music.  Thus, we run a simulation, using the observed recommendation graph, to estimate
what could happen when 500,000 listeners blindly follow
non-personalized music recommendations in passive
listening. Specifically, we simulate a one-hour listening session for each listener by starting
with a track randomly sampled from the top 20\% of most played human-created tracks. The ``listener'' then randomly chooses the next track from the current
track's list of recommendations, until it has listened to 20 additional tracks.

Our results show that 4.40\% of these simulated listening sessions
contain at least one AI track. But more importantly, after listening
to one AI track, the probability that the next song will also be an AI
track is 39.39\%.  {\bf This is alarming, because once a user listens to an
AI-generated track,  the streaming platform is likely to recommend
(and play) additional AI tracks shortly after. }

\subsection{Open Source Music Generators}
\label{app:diffrhythm-acestep}
While we were uploading AI songs ourselves to music streaming platforms, we noticed that all of our DiffRhythm/ACE-step songs were not being labeled as AI by Deezer. This means that our labeling methodology (which uses Deezer's detector) may be underestimating the total amount of AI music. However, given that open source generators still lag behind SOTA in Suno/Udio, it's possible they do not play a large role in the current AI music ecosystem. We do experiments to investigate this here.

First, we trained a highly accurate detector of DiffRhythm and ACE-Step songs using the CNN architecture proposed in~\cite{cnn-reconstruction}. Note that this approach does not generalize to Suno or Udio songs since it relies on having access to a model's latent encoders/decoders, which we only have for DiffRhythm and ACE-Step. Our final model achieved $>$99\% accuracy on both DiffRhythm and ACE-Step songs, with a 0.1\% false positive rate. Finally, we applied the model to a random sample of 10,000 human-made Deezer songs released between October and November 2025.

Our results found that only 0.4\% were identified to be generated via DiffRhythm and ACE-Step. In comparison, our labeled data found that current AI releases (Figure~\ref{fig:ai-music-over-time}) already account for well over 40\% of all weekly releases. This result confirms our suspicion that open source models still lag behind Suno and Udio, and that any unlabeled tracks generated by DiffRhythm or ACE-Step are unlikely to change the overall conclusions of our analysis.

%% file: src/ethics-open-science.tex
\subsection{Ethical Considerations}
\label{app:ethics}
Our study involves potential impact on several primary stakeholder
groups. (1) {\em Streaming platforms} Spotify and Deezer. We used a public
dataset of Spotify metadata in our analyses.  We also used public APIs to
gather data measurements from Spotify's recommendation graph and Deezer's AI
labels. During measurements, we used 6 or fewer machines to crawl
Spotify, and made sure they had publicly identifiable IP addresses so we
could be easily identified and reached if we caused problems. We observed no
server errors or throttling or limiting during data measurements. (2) {\em 11
  Music distributors} that we uploaded music to using standard practice for
indie music artists. We also analyzed their policies on AI music and included
summaries in the paper.
(3) {\em Users of streaming platforms} who have been vocal about
proliferation of AI music and asked for transparency and tools to label AI
music and limit AI slop. (4) {\em Researchers and Practitioners} including
the ML and security community. Groups (3) and (4) were not impacted during
the research process. (5) {\em Human music artists} want to better understand
the impact of AI music and potential slop on their industry and
livelihoods. Our uploads of a small number of AI music tracks received only a
handful of plays, and should not impact human musicians. We also removed all
uploaded AI tracks upon the completion of our study.

\para{Disclosures.} We submitted disclosure to Spotify, informing them that their
public API could put them at risk for data extraction. Spotify replied
thanking us, and stating there is no significant security impact from this
disclosure and they classified our report as ``informative.'' We also
communicated our use of the Deezer API and our results 
directly to the Deezer research team, and their response was quite positive
and cordial. We were careful to not gather any data that could be deemed sensitive or
private. All of our measurements used publicly available APIs, and all data
were public fields that any unauthenticated user could see by browsing Spotify and Deezer
apps. Our interactions with data distributors all followed standard processes
based on their policies and did not disclose any private or sensitive
data.

\para{Copyright.} We did not, at any time, download or store any music from Spotify. 
AI classification was done entirely by matching Spotify tracks to Deezer,
which already labeled AI music.

\para{Justification for research.} This work sheds much needed transparency
on the growth and impact of AI music into popular streaming platforms like
Spotify. Conducting this work is critical to identify potential risks to both
human and AI musicians from AI music slop, and contributes to insights on
potential mitigation mechanisms to limit the harm of AI music slop.